\documentclass[12pt]{article}
\usepackage{multirow}
\usepackage{geometry}
\usepackage{bbold}
\usepackage{amsmath,amssymb,amsfonts}
\usepackage{graphicx}
\usepackage[footnotesize]{caption2}
\usepackage{mathrsfs}
\usepackage{bbm}
\usepackage{braket}
\usepackage{cancel}
\usepackage{color}
\usepackage{enumitem}
\usepackage{cite}
\usepackage{colortbl}
\usepackage{multirow}

\definecolor{Orange}{cmyk}{0,0.61,0.87,0}
\definecolor{JungleGreen}{cmyk}{0.99,0,0.52,0}
\definecolor{OliveGreen}{cmyk}{0.64,0,0.95,0.40}
\definecolor{Brown}{cmyk}{0,0.70,1,0.40}
\definecolor{RoyalBlue}{cmyk}{0.71,0.53,0,0.12}
\definecolor{Gray}{cmyk}{0,0,0,0.40}
\definecolor{LightPink}{cmyk}{0.0,0.25,0,0}
\definecolor{LLightPink}{cmyk}{0.0,0.10,0,0}
\definecolor{LightBlue}{cmyk}{0.25,0,0,0}
\definecolor{LightGray}{cmyk}{0,0,0,0.2}

\usepackage[colorlinks=true, linkcolor=OliveGreen, citecolor=RoyalBlue,
urlcolor=RoyalBlue]{hyperref}

\allowdisplaybreaks[1]

\definecolor{darkblue}{rgb}{0,  0,  .5}
\definecolor{lightgray}{gray}{0.5}

\newcommand{\alphem}{\alpha_{\rm em}}

\newcommand{\eq}[1]{Eq.~\eqref{#1}}
\newcommand{\Figref}[1]{Fig.~\ref{#1}}

\newcommand\sw{{\sin}\theta_{\mathrm{W}}}
\newcommand\cw{{\cos}\theta_{\mathrm{W}}}
\newcommand\swSQ{{\sin}^2\theta_{\mathrm{W}}}
\newcommand\cwSQ{{\cos}^2\theta_{\mathrm{W}}}
\newcommand\swQT{{\sin}^4\theta_{\mathrm{W}}}

\newcommand\sbeta{{\sin}{\beta}}
\newcommand\cbeta{{\cos}{\beta}}

\newcommand{\mc}{{m_{\chi^{\pm}}}}
\newcommand{\mn}{{m_{\chi^{0}}}}
\newcommand{\mH}{{m_{\mathcal{H}}}}
\newcommand{\mcH}{m_{H\pm}}
\newcommand{\mA}{m_{A^0}}

\usepackage{cleveref}

\crefformat{section}{\S#2#1#3} % see manual of cleveref, section 8.2.1
\crefformat{subsection}{\S#2#1#3}
\crefformat{subsubsection}{\S#2#1#3}

\def\beq{\begin{equation}}
\def\eeq{\end{equation}}
\def\bea{\begin{eqnarray} }
\def\eea{ \end{eqnarray} }

\begin{document}

\begin{flushright}
{\tt UMN-TH-4208/23, FTPI-MINN-23/02} \\
{\tt CQUeST-2023-0719 }\\
\end{flushright}

\vspace{0.7cm}
\begin{center}
{\Large\bf 
 Quantifying Limits on CP Violating Phases from EDMs in Supersymmetry
}
\end{center}

\vspace{0.5cm}

\begin{center}{\large
  {\bf Kunio~Kaneta}$^{1}$,
  {\bf Natsumi~Nagata}$^2$,  
  {\bf Keith~A.~Olive}$^3$,
  {\bf Maxim~Pospelov}$^{3,4}$
  and {\bf Liliana~Velasco-Sevilla}$^{5,6}$\\
  }
\end{center}
  
\begin{center}
  {\em $^1$Department of Mathematics, Tokyo Woman?s Christian University, \\ Tokyo 167-8585, Japan} \\[0.2cm]
    {\em $^2$Department of Physics, University of Tokyo,
  Tokyo 113--0033,
  Japan}\\[0.2cm]
    {\em $^3$William I. Fine Theoretical Physics Institute, School of Physics and Astronomy,\\
  University of Minnesota, Minneapolis, MN 55455, USA}\\[0.2cm]
  {\em $^4$ School of Physics and Astronomy, University of Minnesota,\\ Minneapolis, MN 55455, USA}\\[0.2cm]
  {\em $^5$ Center for Quantum Spacetime, Sogang University, Seoul 121-742, South Korea}\\[0.2cm]
  {\em $^6$ Department of Physics, Sogang University, Seoul 121-742, South Korea}
\end{center}
  
\vspace{0.5cm}
\begin{center}
{\bf Abstract}\\
\end{center}
We revisit the calculation of the electron, neutron, and proton electric dipole moments (EDMs) in the constrained minimal supersymmetric standard model (CMSSM). The relatively large mass of the Higgs boson, $m_H \simeq 125$ GeV coupled with the (as yet) lack of discovery of any supersymmetric particle at the LHC, has pushed the supersymmetry breaking scale to several TeV or higher. Though one might expect this decoupling to have relaxed completely any bounds on the two CP violating phases in the CMSSM ($\theta_\mu$ and $\theta_A$), the impressive experimental improvements in the limits on the EDMs (particularly the electron EDM) still allow us to set constraints of order $(0.01 - 0.1)\pi$ on $\theta_A$ and $(0.001 - 0.1)\pi$ on $\theta_\mu$. We also discuss the impact of future improvements in the experimental limits on supersymmetric models.

\newpage

%%%%%%%%%%%%%%%%%%%%%%%%%
\section{Introduction}
%%%%%%%%%%%%%%%%%%%%%%%%%

Supersymmetric extensions of the standard model remain an appealing framework for physics beyond the Standard Model (SM). They go a long way in solving the gauge hierarchy problem ({\em i.e.} the stability of the weak scale) with a minimum amount of fine tuning  \cite{Maiani:1979cx}, particularly if the scale of supersymmetry breaking is $\lesssim \mathcal{O}(1)$ TeV. Supersymmetry is also motivated by grand unification as gauge coupling unification is greatly improved relative to the SM \cite{Ellis:1990zq}. Supersymmetry also provides for a stable electroweak vacuum \cite{Ellis:2000ig} and is consistent with a 125 GeV Higgs bosons mass \cite{mh}. If R-parity is conserved, supersymmetry  of course provides a stable dark matter candidate \cite{gold,ehnos}. 

Searches of supersymmetry (or SUSY), both direct and indirect, have been driving many developments in particle physics, and in particular the collider physics capable of exploring a TeV frontier. The most decisive change in expected scale of supersymmetry breaking occurred with the discovery of the Higgs boson at the LHC \cite{ATLAS:2012yve,CMS:2012qbp}. The mass of the Higgs boson, currently firmly measured to be at 125 GeV, points towards supersymmetry breaking scales in excess of a few TeV and a corresponding heavy scale for superpartner mass spectrum \cite{Bagnaschi:2015eha}. The expected mass scale for squark, sleptons and gauginos is now in tens of TeV \cite{Ellis:2019fwf,Ellis:2022emx}, and is likely outside the direct reach of the LHC, but maybe within a possible range of the next generation of $pp$ colliders. In light of an obvious difficulty with a future collider discovery of superpartners\footnote{Some of superpartners can still be relatively light and accessible by the LHC.}, it is important to evaluate whether the {\em indirect} measurements have a chance of detecting supersymmetric new physics
with the post-Higgs priors on SUSY masses.

There are several important indirect probes of TeV scales physics. These include precision measurements of the muon anomalous magnetic moment, quark flavor transitions, lepton flavor violation and electric dipole moments (EDMs). In this list, only the last two probes are expected to see improvement in sensitivity to new physics by many orders of magnitude \cite{Pospelov:2006jm}, since improving the sensitivity of the muon $g-2$ and quark flavor transitions would require significant improvement in understanding QCD strong interactions. Lepton flavor violation and EDMs, on the other hand, are consistent with zero, and potentially afford orders of magnitude improvement. Indeed, it was noted soon after the Higgs discovery that EDMs have a potential reach to probe superpartner masses in the tens of TeV range \cite{McKeen:2013dma,Moroi:2013sfa}. 

In this paper we give a detailed update to the EDM predictions within the framework of the constrained minimal supersymmetric standard model (CMSSM) \cite{cmssm}. This model assumes that all scalars and all gauginos have a universal mass scale $m_0$ and $m_{1/2}$ respectively at the scale of unification - usually chosen to be the renormalization scale where the electroweak gauge couplings, $g_1$ and $g_2$ are equal. This ansatz makes minimal flavor violation manifest, and leaves EDMs as the best indirect probe of the model. Indeed, it is widely known that within the CMSSM there are two independent flavor-diagonal $CP$-odd phases that can be chosen to be $\theta_\mu$ and $\theta_A$, the phase of the supersymmetric $\mu$-term and the supersymmetry breaking tri-linear term. In the CMSSM, the $\mu$ is fixed (up to a phase) by the minimization of the Higgs potential. The $A$-terms are also taken to be universal with value $A_0$ at the unification scale. The phase of $A$ is not specified by the model.  When the superpartner masses were thought to be at the TeV scale or below, even then existing constraints imposed by the EDMs required small phases (at least of $\theta_\mu$), which was often referred to as ``SUSY $CP$ problem" \cite{Falk:1999tm,Demir:2003js,Lebedev:2004va,Olive:2005ru,Pospelov:2005pr,Pospelov:2005ks,Ellis:2008zy}. These constraints are expected to be relaxed as the supersymmetric mass scales are pushed higher \cite{Lebedev:2002ne,Arbey:2019pdb}.

Parallel to the developments at the LHC, experimental probes of EDMs became much more sensitive. In particular, there has been significant progress in the attempts to measure the EDMs of paramagnetic systems, sensitive primarily to the
electron EDM \cite{ACME:2018yjb,Roussy:2022cmp}. As we will show, these developments are so significant that despite the Higgs-mass-driven requirement of heavy ($\mathcal{O}(10)$ TeV) SUSY masses interesting constraints on all $CP$ phases persist. 

This work is aimed at revisiting the SUSY EDM constraints in light of the new experimental realities. We seek to answer the following questions: 
\begin{enumerate}
    \item Given much heavier superpartner masses, is the SUSY $CP$-problem completely gone? What are the ranges for observable EDMs allowed by the CMSSM for generic values of $CP$ phases?
    
    \item Since 2013, the electron EDM limit was improved by a factor of $O(200)$, while the neutron EDM only by a factor of $\sim 1.5$ \cite{Abel:2020pzs,n2EDM:2021yah}.
    Given new efforts both in $d_e$ and $d_n$, what is the desired improvement in $d_n$ 
    that would allow it to be competitive with $d_e$ within the context of the CMSSM? 
    
    \item In addition to $d_n$ and 
    $d_e$, there exists sensitive probes of EDMs of diamagnetic atoms, and $^{199}$Hg in particular. While the mercury EDM project has finished \cite{Graner:2016ses}, there are certain hopes that the new projects searching for EDMs of atoms enhanced by octupole-deformed nuclei can lead to new levels of sensitivity \cite{Auerbach:1996zd}. What is the desired level of improvement, so that these projects become competitive to $d_e$ in their sensitivity to SUSY? 
    
\end{enumerate}

In what follows, in Section \ref{sec:eff_int},  we will first review the low energy interactions which give rise to the EDMs under consideration. These include the EDMs of paramagnetic atoms and molecules which probe the electron EDM (in Section \ref{paramag}), and nucleons  (in Section \ref{sec:nucleonedm}). In section \ref{susyedm}, we discuss the supersymmetric contributions to EDMs (and color EDMs). In Section \ref{results}, we first introduce the CMSSM parameter space we consider.

Our results are applied to $(m_{1/2}, m_0)$ planes with fixed $A_0$ and $\tan \beta$. 
Though we do not restrict ourselves entirely to regions of parameter space which provide dark matter candidate with a relic density consistent with that measured by CMB experiments \cite{Planck}, our choice of parameter planes is partially motivated by dark matter considerations. We also focus on regions which are consistent with a Higgs mass of 125 GeV. 
As such we focus on four planes: two with $A_0 = 0$ and two with $A_0 = 3 m_0$ for several values of $\tan \beta$. For these choices of parameters, and for fixed values of the phases of $\mu$ and $A_0$, we delineate regions of the parameter space which are constrained by the current limit on the electron EDM, and prospective limits on the neutron and proton EDM. We find that despite the very large mass scales involved ($m_{1/2}, m_0 \gtrsim 10$ TeV), interesting constraints on the phases may still be derived, largely due to the vast improvement in the limit on the electron EDM. We also show that for the neutron EDM to be competitive with the electron EDM, improvements by roughly a factor of $10^3$ in the limits to the neutron EDM are needed.

%%%%%%%%%%%%%%%%%%%%%%%%%%%%%%%%%%%%%%%%%%%%
\section{Low-energy effective interactions}
\label{sec:eff_int}
%%%%%%%%%%%%%%%%%%%%%%%%%%%%%%%%%%%%%%%%%%%%

The CP-violating flavor-conserving interactions at low energies can be described by the following effective CP-odd operators:\footnote{For more detailed discussions on the material given in this section, see, \textit{e.g.}, Refs.~\cite{Khriplovich:1997ga, Pospelov:2005pr, Chupp:2017rkp}. }
\begin{align}
  \mathcal{L}_{\mathrm{eff}} &= \frac{g_s^2}{32\pi^2} \theta_G G^a_{\mu\nu} \widetilde{G}^{a \mu\nu} + \sum_{q } m_q  \theta_q \bar{q} i \gamma_5 q \nonumber \\ 
  &- \frac{i}{2} \sum_{f = q,\ell} d_f \bar{f} \sigma^{\mu\nu} \gamma_5 f F_{\mu\nu} 
  - \frac{i}{2} \sum_{q} \tilde{d}_q  g_s \bar{q} \sigma^{\mu\nu} \gamma_5 T^a q \,  G^a_{\mu\nu} \nonumber \\ 
  &+\frac{1}{3}w f_{abc} G^a_{\mu\nu} \widetilde{G}^{b \nu\rho} G^{c\,\mu}_{\rho}
  + 
  \sum_{f, f^\prime = q,\ell} C_{f f^\prime} \bar{f} f \bar{f}^\prime i\gamma_5 f^\prime 
  + \dots ~,
  \label{eq:leff}
\end{align}
where $\widetilde{G}^a_{\mu\nu}$ and $\widetilde{F}_{\mu\nu}$ are the dual field-strength tensors of the gluon and electromagnetic fields, $G^a_{\mu\nu} \equiv \frac{1}{2}\epsilon_{\mu\nu\rho\sigma} G^{a\rho\sigma}$ and $F_{\mu\nu} \equiv \frac{1}{2}\epsilon_{\mu\nu\rho\sigma} F^{\rho\sigma}$, respectively, $\epsilon^{\mu\nu\rho\sigma}$ is the totally antisymmetric tensor with $\epsilon^{0123} = +1$, $g_s$ is the strong gauge coupling constant, $m_q$ denote quark masses, and $T^a$ and $f_{abc}$ are the generators and structure constant of the SU(3) gauge group, respectively. The effective theory described by these operators is valid below the electroweak scale down to the hadronic scale, $\sim 1$~GeV. The sum with respect to $q$ and $\ell$ is taken over the active quark and lepton flavors for each scale, respectively. The first two terms in Eq.~\eqref{eq:leff} are converted into each other through chiral rotations, and thus only a linear combination of the coefficients of these terms, $\bar{\theta} \equiv \theta_G + \sum_q \theta_q$, is the physical quantity. In this work, we assume that this physical $\bar{\theta}$ term is suppressed by the Peccei-Quinn mechanism~\cite{Peccei:1977hh}, and thus we do not consider the first two terms in Eq.~\eqref{eq:leff} in the following discussions.\footnote{We note, however, that in the presence of the CP-violating interactions, the minimum of the axion potential~\cite{Wilczek:1977pj} deviates from that in the vacuum without these interactions and thus the effective $\bar{\theta}$ term becomes non-vanishing~\cite{Bigi:1991rh, Pospelov:1997uv}. We take this effect into account in the following calculation.  } The next-to-lowest dimensional operators are of dimension-five, corresponding to the third and fourth terms in Eq.~\eqref{eq:leff}; these operators are called the electric dipole moments (EDMs) and chromo-electric dipole moments (CEDMs), respectively. However, above the electroweak scale these operators are matched onto the dimension-six operators with an insertion of the Higgs field, required by their chiral structure. We, therefore, also consider the following dimension-six operators for consistency. The Weinberg operator~\cite{Weinberg:1989dx} is the fifth term in Eq.~(\ref{eq:leff}) and the sixth term represents a part of the CP-odd four fermion operators. Among the possible Lorentz structures of the four fermion operators, we only show one class of operators relevant to our discussions and the rest of them are included in the dots in Eq.~\eqref{eq:leff}; for a complete list of the CP-odd dimension-six operators, see Ref.~\cite{Khatsimovsky:1987fr}. As discussed in Refs.~\cite{Lebedev:2002ne, Demir:2003js}, in supersymmetric models, the contribution of the four-fermion operators can be significant for large $\tan\beta$. 

The dimension-five and -six operators in Eq.~\eqref{eq:leff} mix with each other under the renormalization group (RG) flow~\cite{Braaten:1990gq}. We summarize the RG effect on these operators in Appendix~\ref{sec:rge}. 

Below the hadronic scale, the interactions in Eq.~\eqref{eq:leff} that contain quark and gluon fields give rise to the effective operators consisting of nucleon and pion fields. Let us write down the operators relevant to the following discussions. First, we have the EDMs of nucleons and leptons, 
\begin{equation}
  \mathcal{L}_{\mathrm{EDM}} = 
  - \frac{i}{2} \sum_{\ell = e, \mu} d_\ell \, \bar{\ell} \sigma^{\mu\nu} \gamma_5 \ell F_{\mu\nu}
  - \frac{i}{2} \sum_{N = p,n} d_N \bar{N} \sigma^{\mu\nu} \gamma_5 N F_{\mu\nu}  ~. 
  \label{eq:ledm}
\end{equation}
The EDMs of leptons are the same as those in Eq.~\eqref{eq:leff}, while those of nucleons are given as functions of the coefficients of the CP-violating operators in Eq.~\eqref{eq:leff}; we show the relevant expressions in Sec.~\ref{sec:nucleonedm}. The second class of operators consist of the CP-odd pion-nucleon couplings:
\begin{align}
  \mathcal{L}_{\pi NN} &= \bar{g}_{\pi NN}^{(0)} \bar{N} \tau^a N \pi^a
  +\bar{g}_{\pi NN}^{(1)} \bar{N} N \pi^0
  + \bar{g}_{\pi NN}^{(2)} (\bar{N} \tau^a N \pi^a - 3 \bar{N} \tau^3 N \pi^0) ~, 
\end{align} 
where $\tau^a$ are the isospin Pauli matrices.
Finally, we have the nucleon-electron four-fermion operators\footnote{We define $\gamma_5 \equiv i \gamma^0 \gamma^1 \gamma^2 \gamma^3$ in accordance with Ref.~\cite{Demir:2003js}. Note that this has the opposite sign to that used in, \textit{e.g.}, Refs.~\cite{Khriplovich:1997ga, 2012PhRvA..85b9901D}. }
\begin{align}
  \mathcal{L}_{eN} &= C_S^{(0)}\, \bar{e} i \gamma_5 e \bar{N} N + C_P^{(0)}\, \bar{e}e \bar{N} i \gamma_5 N + C_T^{(0)} \, \epsilon_{\mu\nu\rho\sigma} \bar{e} \sigma^{\mu\nu} e \bar{N} \sigma^{\rho\sigma} N \nonumber \\ 
  &+ C_S^{(1)} \, \bar{e} i \gamma_5 e \bar{N} \tau^3 N + C_P^{(1)}\, \bar{e}e \bar{N} i \gamma_5 \tau^3 N + C_T^{(1)} \, \epsilon_{\mu\nu\rho\sigma} \bar{e} \sigma^{\mu\nu} e \bar{N} \sigma^{\rho\sigma} \tau^3 N ~.
  \label{eq:len}
\end{align}
These interactions are induced by the semi-leptonic four-fermion operators in Eq.~\eqref{eq:leff}.

The above interactions then induce the EDMs of nucleons, atoms, and molecules. We summarize the relation between these EDMs and the interactions in Eq.~\eqref{eq:leff} in what follows. 

%%%%%%%%%%%%%%%%%%%%%%%%%%%%%%%%%%%%%%%%%%%%%%
\subsection{Paramagnetic atoms and molecules}
\label{paramag}
%%%%%%%%%%%%%%%%%%%%%%%%%%%%%%%%%%%%%%%%%%%%%%

Paramagnetic atoms and molecules, which have unpaired electrons, offer a sensitive probe to the electron EDM and the electron-nucleon four-fermion interactions. The EDM of a paramagnetic atom induced by these interactions has the form 
\begin{equation}
  d_A = \alpha_{A, d_e} d_e \, + \alpha_{A, C_S} \cdot \frac{\sqrt{2}}{G_F} C_S ~, 
\end{equation}
where $G_F$ is the Fermi constant, $\alpha_{A, d_e} $ and $\alpha_{A, C_S}$ are parameters that depend on each atom, and 
\begin{equation}
  C_S \equiv C_S^{(0)} + \biggl(\frac{Z-N}{Z+N}\biggr) C_S^{(1)} ~,
  \label{eq:csdef}
\end{equation}
with $Z$ and $N$ the numbers of protons and neutrons in the atom $A$, respectively. In Refs.~\cite{Demir:2003js, Olive:2005ru}, $\alpha_{{\rm Tl}, d_e} = -585$ and $\sqrt{2}\alpha_{{\rm Tl}, C_S}/G_F = -43~e \cdot \mathrm{GeV}$ were used for the thallium atom. 
For paramagnetic molecules, on the other hand, the experimental limits are usually given in terms of the frequency shift under the electric field. This is, again, expressed in the form 
\begin{equation}
  \Delta \omega_M = \beta_{M, d_e} d_e + \beta_{M, C_S}\cdot \frac{\sqrt{2}}{G_F} C_S ~. 
\end{equation} 
The parameters $\beta_{M, d_e}$ and $\beta_{M, C_S}$ are obtained from molecular-structure computation. 

To obtain limits on $d_e$ and $C_S$, it is useful to introduce the ``equivalent electron EDM'' defined by~\cite{Pospelov:2013sca}
\begin{equation}
  d_e^{\mathrm{equiv}} \equiv d_e + r_A \cdot \frac{\sqrt{2}}{G_F} C_S ~,
  \label{eq:deequiv} 
\end{equation}
where the coefficient $r_A$ depends on the atom (molecule) used to impose a limit on $d_A$ ($\Delta \omega_M$). For a paramagnetic atom, $r_A = \alpha_{A, C_S}/ \alpha_{A, d_e}$, while for a molecule, $r_A = \beta_{M, C_S}  / \beta_{M, d_e} $. Experimental limits are often given as an upper limit on $|d_e|$ with $C_S = 0$ assumed; in this case, we can interpret this limit by just replacing $|d_e|$ by $|d_e^{\mathrm{equiv}}|$ for a non-zero $C_S$.

As seen in Eq.~\eqref{eq:csdef}, paramagnetic systems are sensitive to $C_S^{(0)}$ and $C_S^{(1)}$ in Eq.~\eqref{eq:len}. These couplings are related to the semi-leptonic four-fermion couplings as 
\begin{align}
  C_S^{(0)} &= \sum_q \frac{C_{qe}}{ 2m_q}  \left(  m_p f^p_{T_q} + m_n f^n_{T_q}\right) ~, \nonumber \\ 
  C_S^{(1)} &= \sum_q \frac{C_{qe}}{2 m_q} \left(  m_p f^p_{T_q} - m_n f^n_{T_q}\right)~,
  \label{eq:cs0andcs1}
\end{align}
where $f^N_{T_q} \equiv \langle N | m_q \bar{q} q | N\rangle/m_N$ ($N = p,n$) are the nucleon matrix elements of the scalar-type quark operators, with $m_N$ the nucleon mass. For light quarks, the matrix elements are obtained from lattice simulations or phenomenological estimations based on the baryon chiral perturbation theory. In the present work, we use the values obtained by a recent compilation~\cite{Ellis:2018dmb}: $f^p_{T_u} = 0.018(5)$, $f^n_{T_u} = 0.013(3)$, $f^p_{T_d} = 0.027(7)$, $f^n_{T_d} = 0.040(10)$, $f^p_{T_s} = f^n_{T_s} = 0.037(17)$. Heavy quarks contribute to the nucleon matrix elements via gluons at the loop level, whose nucleon matrix element can be obtained from the light-quark matrix elements through the trace anomaly of the energy-momentum tensor in QCD~\cite{Shifman:1978zn}. At the leading order in $\alpha_s \equiv g_s^2/(4\pi)$, all of the heavy quark contributions are the same, since the scalar-type gluon operator $\alpha_s G^a_{\mu\nu} G^{a \mu\nu}$ is invariant under the RG flow. Small difference among them arises at higher orders in $\alpha_s$, and we find $f^p_{T_c} = f^n_{T_c} = 0.078(2)$, $f^p_{T_b} = 0.072(2)$, $f^n_{T_b} = 0.071(2)$, $f_{T_t}^p = 0.069(1)$, $f_{T_t}^n = 0.068 (2)$~\cite{Ellis:2018dmb}.

In the present case, we can safely neglect some of the terms in Eq.~\eqref{eq:cs0andcs1}. First, the isospin violation in the nucleon matrix elements is fairly small, which makes $|C_S^{(1)}| \ll |C_S^{(0)}|$. Given that the contribution of $C_S^{(1)}$ to Eq.~\eqref{eq:csdef} is further suppressed by the factor $(Z-N)/(Z+N)$, we can ignore $C_S^{(1)}$. Second, as discussed in Refs.~\cite{Lebedev:2002ne, Demir:2003js}, in the MSSM $C_S$ can be sizable only for large $\tan \beta$. In this case, the down-type quarks predominantly contribute to Eq.~\eqref{eq:cs0andcs1}. We therefore have an approximated expression for $C_S$ as\footnote{To compare this with the previous results~\cite{Lebedev:2002ne, Demir:2003js, Olive:2005ru}, we recalculate the values of the input parameters used in these previous works: $\kappa \equiv \langle N| m_s \bar{s}s |N\rangle/(220~\mathrm{MeV})$ and $\Sigma_{\pi N} \equiv (m_u + m_d) \langle N | \bar{u}u + \bar{d}d |N\rangle/2$. Using the results in Ref.~\cite{Ellis:2018dmb}, we have $\kappa = \sigma_s/(220~\mathrm{MeV}) \simeq 0.16(7)$ and $\Sigma_{\pi N} = 46(11)$~MeV. As we will see, the value of $\kappa$ obtained here is much smaller than those used in the previous studies, which results in a smaller coefficient for $C_{se}/m_s$ in Eq.~\eqref{eq:csapprox}. } 
\begin{equation}
  C_S \simeq \frac{C_{de}}{m_d} \cdot 31~\mathrm{MeV} + \frac{C_{se}}{m_s} \cdot 35~\mathrm{MeV} + \frac{C_{be}}{m_b} \cdot 67~\mathrm{MeV} ~.
  \label{eq:csapprox}
\end{equation}
We note that in our setup, $C_{qe} \propto m_q$, which cancel the down-quark masses in the denominators in the above expression.

%%%%%%%%%%%%%%%%%%%%%%%%%%%%%%%%%%%%%%%%%%%%%%%%%%%%%%%%%%%%%%%%%%%%%%%%%%%%%%%5
\begin{table}[t]
  \centering
  \caption{Limits on $|d_e|$ for $C_S = 0$ and the coefficient $r_A$ for each paramagnetic atom/molecule.  }
  \vspace{3mm}
  \begin{tabular}{lll}
    \hline\hline
     & Limit on $|d_e|$ ($e\cdot \mathrm{cm}$) & $r_A$ ($e\cdot \mathrm{cm}$)\\ 
    \hline
    \rowcolor{LightGray}
    Atom &&\\
    \hline 
     Tl & $1.6 \times 10^{-27}$ (90\% CL)~\cite{Regan:2002ta} &  $1.2 \times 10^{-20}$~\cite{2012PhRvA..85b9901D} \\
     % Cs &  & \\ 
     \hline
     \rowcolor{LightGray}
     Molecule &&\\
     \hline 
     ThO & $1.1 \times 10^{-29}$ (90\% CL)~\cite{ACME:2018yjb} & $1.3 \times 10^{-20}$~\cite{2012PhRvA..85b9901D}\\ 
     HfF$^+$ & $4.1 \times 10^{-30}$ (90\% CL)~\cite{Roussy:2022cmp} &  $9.2 \times 10^{-21}$~\cite{Roussy:2022cmp} \\
     YbF & $1.05 \times 10^{-27}$ (90\% CL)~\cite{Hudson:2011zz} & $8.8 \times 10^{-21}$~\cite{2012PhRvA..85b9901D} \\ 
    \hline\hline
  \end{tabular}
  \label{tab:paramag}
\end{table}
%%%%%%%%%%%%%%%%%%%%%%%%%%%%%%%%%%%%%%%%%%%%%%%%%%%%%%%%%%%%%%%%%%%%%%%%%%%%%%%%

In Table~\ref{tab:paramag}, we list the present limits on $|d_e|$ set by the measurements of paramagnetic atoms and molecules for $C_S = 0$, as well as the value of $r_A$ for each system. The corresponding limits for $C_S \ne 0$ can be obtained by replacing $|d_e|$ with $|d_e^{\mathrm{equiv}}|$ defined in Eq.~\eqref{eq:deequiv}. At the time of Refs.~\cite{Demir:2003js, Olive:2005ru}, the most stringent bound on $|d_e|$ was given by the measurement of thallium; $|d_{\mathrm{Tl}}| < 9 \times 10^{-25}~e \cdot \mathrm{cm}$, which corresponds to $|d_{e}| < 1.6 \times 10^{-27}~e \cdot \mathrm{cm}$ for $C_S = 0$~\cite{Regan:2002ta}. 
Presently, the JILA electron EDM experiment using HfF$^+$ molecular ions imposes the strongest constraint on the electron EDM \cite{Roussy:2022cmp}: 
\beq
|d_e| < 4.1 \times 10^{-30}~e \cdot {\rm cm}
\eeq
stronger than the Tl bound by about three orders of magnitude!

In addition, there are many proposals for future measurements of the EDMs of paramagnetic atoms and molecules, which are expected to improve the sensitivity by several orders of magnitude; see Ref.~\cite{Alarcon:2022ero} for a recent review. Future experiments rely on the advancement in atomic, molecular, and optical (AMO) techniques, such as laser cooling, trapping, and quantum control of molecules; with these techniques, the experimental sensitivity is expected to improved by several orders of magnitude in $\sim 10$~years~\cite{Alarcon:2022ero}. 

%%%%%%%%%%%%%%%%%%%%%%%%
\subsection{Nucleon}
\label{sec:nucleonedm} 
%%%%%%%%%%%%%%%%%%%%%%%

Next, we discuss the relation between the nucleon EDMs in Eq.~\eqref{eq:ledm} and the CP-odd quark and gluon operators in Eq.~\eqref{eq:leff}. The contribution of quark EDMs and CEDMs to nucleon EDMs can be computed on an equal footing by means of the QCD sum-rule technique~\cite{Shifman:1978bx}, as described in Refs.~\cite{Pospelov:1999ha, Pospelov:1999mv, Pospelov:2000bw, Hisano:2012sc}. As discussed in Ref.~\cite{Pospelov:2005pr}, the dominant contribution of the quark EDMs and CEDMs to the neutron EDM has the form\footnote{Sub-leading terms include the condensates $g_s \langle \bar{q} G^a_{\mu\nu} T^a q \rangle_F$ and $g_s \langle \bar{q} \widetilde{G}^a_{\mu\nu} T^A  \gamma_5 q \rangle_F$~\cite{Pospelov:2000bw}, which are numerically less significant compared with the quark condensate electromagnetic susceptibility $\chi_q$.  } 
\begin{equation}
  d_n = C_n \left[ 4 \chi_d m_d  P_d - \chi_u m_u P_u   + 4 d_d - d_u  \right] ~,
  \label{eq:dngen}
\end{equation}
where $C_n$ is an overall normalization factor that depends on the Borel mass $M$, the quark condensate $\langle \bar{q} q \rangle$, and the size of the coupling between the one-particle neutron state and the interpolating neutron current, $\lambda_n$; $P_{u (d)}$ is defined by 
\begin{equation}
  m_{u (d)}P_{u (d)} \equiv m_* \bar{\theta} + \frac{m_* M_0^2}{2} \biggl(\frac{\tilde{d}_{u(d)} - \tilde{d}_{d(u)} }{m_{d (u)}} + \frac{ \tilde{d}_{u(d)} - \tilde{d}_s}{m_s}\biggr) ~,
\end{equation}
with 
\begin{equation}
  m_* \equiv \frac{m_u m_d m_s}{m_u m_d + m_d m_s + m_u m_s} ~,
\end{equation}
and \footnote{Please note the difference between ratio of the quark-gluon dipole condensate to the scalar quark condensate, $M_0$, and the universal supersymmetry breaking scalar mass, $m_0$.}
\begin{equation}
 M_0^2 \langle \bar{q} q \rangle  \equiv - g_s \langle \bar{q}  \sigma^{\mu\nu} G_{\mu\nu}^a T^a q \rangle  ~. 
  \label{eq:m02def}
\end{equation}
The parameters $\chi_q$ in Eq.~\eqref{eq:dngen} are given by the quark condensate electromagnetic susceptibility
\begin{equation}
  \langle \bar{q} \sigma_{\mu\nu} q \rangle_F = \chi_q F_{\mu\nu} \langle \bar{q}q \rangle ~.
\end{equation}
We note that when we obtain the formula~\eqref{eq:dngen}, we choose a specific form of the interpolating neutron current to suppress the next-to-leading order terms, which depend logarithmically on an infrared cutoff~\cite{Pospelov:1999ha, Pospelov:1999mv, Pospelov:2000bw}. 

As mentioned in Sec.~\ref{sec:eff_int}, we assume the Peccei-Quinn mechanism to suppress the $\bar{\theta}$ term. In the presence of the quark CEDMs, however, the minimum of the axion potential shifts from the point $\bar{\theta} = 0$ so that an effective $\bar{\theta}$ term is induced~\cite{Bigi:1991rh, Pospelov:1997uv}: 
\begin{equation}
  \bar{\theta}_{\mathrm{ind}} = \frac{M_0^2}{2} \sum_{q = u,d,s} \frac{\tilde{d}_q}{m_q} ~.
\end{equation}
By using this induced $\bar{\theta}$ term in Eq.~\eqref{eq:dngen}, we obtain 
\begin{equation}
  d_n = C_n \left[ \frac{M_0^2}{2}  \left(4 \chi_d \tilde{d}_d - \chi_u \tilde{d}_u \right) + 4 d_d - d_u \right] ~.
  \label{eq:dnpq}
\end{equation}

For the parameter $M_0^2$ in Eq.~\eqref{eq:m02def}, the previous calculations~\cite{Pospelov:2000bw, Pospelov:2005pr, Hisano:2012sc} used the value determined with the QCD sum rules~\cite{Belyaev:1982sa}: $M_0^2 = 0.8(2)~\mathrm{GeV}^2$. This quantity is computed also with lattice simulations: $M_0^2 \simeq 2.5~\mathrm{GeV}^2$ in Ref.~\cite{Doi:2002wk} and $M_0^2 = 0.98(2)~\mathrm{GeV}^2$ in Ref.~\cite{Chiu:2003iw}. As pointed out in Ref.~\cite{Gubler:2018ctz}, however, these lattice simulations do not take account of operator mixing, which may cause systematic uncertainty in the above estimations. Considering this, we use in this paper the sum-rule value, $M_0^2 = 0.8(2)~\mathrm{GeV}^2$, as in previous works~\cite{Pospelov:2000bw, Pospelov:2005pr, Hisano:2012sc}. 

The electromagnetic susceptibility of the condensate was evaluated with QCD sum rules in Ref.~\cite{Belyaev:1984ic} with the assumption $\chi_q = e Q_q \chi$~\cite{Ioffe:1983ju}, where $Q_q$ denotes the quark electric charge: $\chi = - 5.7 (6)~\mathrm{GeV}^{-2}$ in the zero momentum limit. This value was used in the previous works~\cite{Pospelov:2000bw, Pospelov:2005pr, Hisano:2012sc}. At the renormalization scale of $\simeq m_N$, the same calculation results in $\chi \simeq -4.4~\mathrm{GeV}^{-2}$~\cite{Balitsky:1985ag}. A more recent QCD sum-rule calculation~\cite{Ball:2002ps}, which uses new data for the resonances and takes account of QCD radiative corrections, gives $\chi = -(3.15\pm 0.3)~\mathrm{GeV}^{-2}$ at the renormalization scale of 1~GeV. On the other hand, a recent lattice simulation determines $\chi_q$ at the renormalization scale of 2~GeV as $\chi_u/(e Q_u) = - 2.08(8)~\mathrm{GeV}^{-2}$ and $\chi_d/(e Q_d) = - 2.02(9)~\mathrm{GeV}^{-2}$~\cite{Bali:2012jv}.\footnote{The values at the scale of 1~GeV can be obtained by multiplying the 2-GeV values by a factor of $1.49 (7)$~\cite{Bali:2012jv}. } We use these values in our  calculations. We note, however, that arguments motivated by the large $N_c$ approximation may suggest that the {\em product} of two condensates is in fact fixed, $ M_0^2 \chi \simeq 6$ \cite{Cata:2009fd}. This will result in a considerably larger $\tilde d_i$ contributions (by a factor of $\sim 3$), and closer to the original estimates in~\cite{Pospelov:2000bw, Pospelov:2005pr, Hisano:2012sc}. We therefore conclude that perhaps a {\em simultaneous} lattice determination of $M_0^2\chi$ is needed to clarify this issue.

The determination of the overall constant $C_n$ suffers from the uncertainty originating from the choice of the Borel mass and the coupling $\lambda_n$. The latter uncertainty may be reduced~\cite{Hisano:2012sc} by using lattice results for $\lambda_n$, which can be obtained from the parameters $\alpha$ and $\beta$ in, \textit{e.g.}, Ref.~\cite{Yoo:2021gql}. Instead, in the present work, we determine the value of $C_n$ by requiring that the quark EDM contribution to the neutron EDM agrees with the lattice result. This contribution can be obtained from the lattice calculations of nucleon tensor charges~\cite{Gupta:2018qil, Gupta:2018lvp, Alexandrou:2019brg}. We use the result obtained in Ref.~\cite{Alexandrou:2019brg} and fix $C_n$ such that the coefficient of the $d$-quark EDM contribution matches the corresponding tensor charge, $g_T^d = 0.729(22)$ for neutron: $4 C_n = g_T^d$, \textit{i.e.}, $C_n \simeq 0.182$. This tensor charge is renormalized at the scale of 2~GeV. With this choice, the coefficient of the $u$-quark contribution in Eq.~\eqref{eq:dnpq} is $- 0.182$, which agrees with the result in Ref.~\cite{Alexandrou:2019brg}, $g_T^u = -0.2075(75)$, within an accuracy of $\sim 10$\%. The $s$-quark EDM contribution is also evaluated in Ref.~\cite{Alexandrou:2019brg}: $g_T^s = -0.00268(58)$. This is smaller than $g_T^d$ by a factor of $\sim 300$, so we can safely neglect the $s$-quark EDM contribution in the calculations below.

By substituting all of the above numerical parameters into Eq.~\eqref{eq:dnpq}, we finally obtain 
\begin{align}
  d_n &= 0.73 \, d_d - 0.18 \, d_u + e (0.20\, \tilde{d}_d + 0.10 \,\tilde{d}_u) ~, \label{eq:dnnum}
  \\[3pt] 
  d_p &= 0.73 \, d_u - 0.18 \, d_d - e (0.40\, \tilde{d}_u + 0.049 \,\tilde{d}_d) ~,
\end{align}
where we have used the isospin relation to obtain the proton EDM. All of the couplings in the above equations should be evaluated at the scale of 2~GeV. Compared to the previous calculation in Ref.~\cite{Pospelov:2005pr}, the quark EDM contribution is smaller by a factor of $2$, which is due to a smaller value of $C_n$. The quark CEDM contribution receives an additional suppression factor of a few, which originates from the difference in $\chi_q$. The quark EDM contribution given in Ref.~\cite{Fuyuto:2013gla} agrees with Eq.~\eqref{eq:dnnum}, but the quark CEDM contribution differs by a factor of 3, which is again attributed to the difference in $\chi_q$. 

The estimation of the contribution of the Weinberg operator to the nucleon EDMs suffers from larger uncertainty than that of the EDMs and CEDMs. In Refs.~\cite{Demir:2002gg, Haisch:2019bml}, this contribution is estimated using the QCD sum-rule technique on the assumption that it is predominantly given by the chiral rotation of the nucleon magnetic moment, $\mu_N$, induced by the Weinberg operator:\footnote{This assumption turns out to be plausible as the 1PI contribution is estimated to be subdominant~\cite{Yamanaka:2020kjo}. }
\begin{align}
  d_N|_w = - \mu_N \frac{3 g_s M_0^2}{32 \pi^2} \ln \biggl(\frac{M^2}{\mu_{\mathrm{IR}}^2}\biggr)  \cdot w ~,
\end{align}
where $M$ is the Borel mass and $\mu_{\mathrm{IR}}$ is an infrared (IR) cutoff. The presence of the IR cutoff in this expression is due to a logarithmic IR divergence in the operator product expansion (OPE) of the nucleon current correlator, implying the failure of the OPE. In the following numerical calculation, we set $M/\mu_{\mathrm{IR}} = 2$ and $g_s = 2.1$, keeping in mind that this estimation may suffer from an $\mathcal{O}(1)$ uncertainty due to this problem.  By using $\mu_n = -1.91 \mu_N$ and $\mu_p = 2.79 \mu_N$ with $\mu_N = e/(2m_p)$~\cite{ParticleDataGroup:2020ssz}, we obtain 
\begin{align}
  d_n|_w &\simeq e \cdot (23~{\rm MeV}) \cdot w 
  ~, \\ 
  d_p|_w &= - e \cdot (33~{\rm MeV}) \cdot w  ~.
\end{align}

Currently, the best limit on the neutron EDM is set by the experiment performed at the Paul Scherrer Institute (PSI)~\cite{Abel:2020pzs}:
\begin{equation}
    |d_n| < 1.8 \times 10^{-26}~e \cdot {\rm cm} ~.
\end{equation}
The n2EDM experiment at PSI is expected to improve the sensitivity by more than an order of magnitude, $\sigma(d_n) = 1.1 \times 10^{-27}~e \cdot {\rm cm}$~\cite{n2EDM:2021yah}. The TUCAN experiment at the TRIUMF also aims for a sensitivity of this size~\cite{Martin:2020lbx}. The statistical sensitivity of the PanEDM experiment at the Institut Laue Langevin (ILL) is expected to be $\sigma(d_n) = 3.8 \times 10^{-27}~e \cdot {\rm cm}$ in Phase I, which will be improved to $\sigma(d_n) = 7.9 \times 10^{-28}~e \cdot {\rm cm}$ in Phase II~\cite{Wurm:2019yfj}. At the Los Alamos National Laboratory (LANL), an experiment using the LANL ultracold neutron source is planned, which is expected to offer a sensitivity of $\sigma(d_n) = 3 \times 10^{-27}~e \cdot {\rm cm}$~\cite{Ito:2017ywc}. The nEDM@SNS experiment at the Oak Ridge National Laboratory aims at even better sensitivity, $\sigma(d_n) = (2-3) \times 10^{-28}~e \cdot {\rm cm}$~\cite{nEDM:2019qgk}. 

For the present limit on the proton EDM, $|d_p| < 2.1 \times 10^{-25}~e \cdot \mathrm{cm}$ is adopted in Ref.~\cite{ParticleDataGroup:2020ssz}, which is indirectly obtained from the limit on the EDM of ${}^{199}\mathrm{Hg}$~\cite{Sahoo:2016zvr}. This limit can be improved by orders of magnitude with a storage ring---the expected sensitivity is as low as $\sim 10^{-29}~e \cdot \mathrm{cm}$~\cite{Alarcon:2022ero}. 

%%%%%%%%%%%%%%%%%%%%%
\begin{table}[!ht]
  \centering
  \caption{The current limit and future sensitivity of nucleon EDMs. }
\vspace{3mm}
  \begin{tabular}{lll}
    \hline\hline
       & Current limit ($e\cdot \mathrm{cm}$; 90\% CL) & Future sensitivity ($e\cdot \mathrm{cm}$) \\ 
    \hline 
   Neutron & $1.8 \times 10^{-26}$~\cite{Abel:2020pzs} & $\sim (2-3) \times 10^{-28}$~\cite{nEDM:2019qgk} \\
     \hline
   Proton & $2.1 \times 10^{-25}$~\cite{Sahoo:2016zvr} & $\sim 10^{-29}$~\cite{Alarcon:2022ero} \\
    \hline\hline
  \end{tabular}
  \label{tab:nucleon}
\end{table}
%%%%%%%%%%%%%%%%%%%%%%%%%%%%%%%%%%%%%%%%%%%%%%%%%%%%%%%%%%%%%%%%%%%%%%%%%%%%%%%%

In Table~\ref{tab:nucleon}, we summarize the current limit and future sensitivity of  nucleon EDMs.

%%%%%%%%%%%%%%%%%%%%%%%%%%%%%%%%%%%%%%%%%%%%%%%%%%%%%%%%%%%%%%%%%%%%%%%
\section{Supersymmetric contribution to the (C)EDMs}
\label{susyedm}
%%%%%%%%%%%%%%%%%%%%%%%%%%%%%%%%%%%%%%%%%%%%%%%%%%%%%%%%%%%%%%%%%%%%%%%

In this Section, we briefly summarize the well known one-loop SUSY contributions to the EDMs and CEDMs \cite{Ibrahim:1997gj,Ibrahim:1998je}.
In Appendix \ref{sec:2lsusy} we present explicit expressions from two-loop diagrams and a comparison of the two-loop and  one-loop contributions. 
\footnote{Our notation is also specified in Appendix \ref{sec:2lsusy}.}
We have updated the code used in \cite{Demir:2003js,Lebedev:2004va,Olive:2005ru}, taking into account all the updates discussed in Section \ref{sec:eff_int} and the 2-loop contributions described in  Appendix \ref{sec:2lsusy}.

\subsection{One loop contributions to the EDMs}

The EDMs of charged leptons and light quarks ($u,d,s$) receive gaugino/higgsino and sfermion contributions through the fermion-sfermion-gaugino/higgsino interactions.
We categorize these contributions according to which gaugino/higgsino is exchanged at one loop.
Note that our convention for the relevant $CP$ phases are those in the $\mu$- and $A$-terms, namely, $\theta_\mu={\rm arg}(\mu)$ and $\theta_A={\rm arg}(A)$, where the phase in gaugino mass is rotated out, and the flavor dependence is omitted since we consider the CMSSM.

\subsubsection{Gluino contribution}
The gluino contribution to the quark EDMs, denoted by $d_q^{\tilde g}$, is given by
\begin{align}
    \frac{d_q^{\tilde g}}{e} &=
    -\frac{2\alpha_s}{3\pi}\sum_{k=1}^2{\rm Im}[S^*_{q2k}S_{q1k}]\frac{m_{\tilde g}}{m^2_{\tilde{q}_k}}Q_{\tilde q}B(m^2_{\tilde g}/m^2_{\tilde{q}_k}),
\end{align}
where $Q_f$ is the electromagnetic charge of fermion or sfermion, $f$.

\subsubsection{Neutralino contribution}
The neutralino contribution to the fermion EDMs, $d^{\widetilde{\chi}^0}_f$, is given by
\begin{align}
\label{eq:neutf}
    \frac{d^{\widetilde{\chi}^0}_f}{e} &=
    \frac{\alpha_{\rm em}}{4\pi\sin^2\theta_W}\sum_{k=1}^2\sum_{i=1}^4{\rm Im}[\eta_{fik}]\frac{m_{\tilde{\chi}^0_i}}{m^2_{\tilde{f}_k}}Q_{\tilde f}B(m^2_{\tilde{\chi}^0_i}/m^2_{\tilde{f}_k}),
\end{align}
where $\alpha_{\rm em}\equiv e^2/4\pi$, and
\begin{align}
    \eta_{fik} &=
    \left[-\sqrt{2}\left\{\tan\theta_W(Q_f-T_{3f})N^*_{1i}+T_{3f}N^*_{2i}\right\}S_{f1k}+\kappa_f N^*_{bi}S_{f2k}\right]\nonumber\\
    &\times\left(\sqrt{2}\tan\theta_W Q_f N^*_{1i}S^*_{f2k}-\kappa_f N^*_{bi}S^*_{f1k}\right),\\
    \kappa_u &= \frac{m_u}{\sqrt{2}M_W\sin\beta},\;\;\;
    \kappa_{d,e} = \frac{m_{d,e}}{\sqrt{2}M_W\cos\beta},
\end{align}
with $b$ referring to the higgsino contribution where $b=3$ for $f=d, e$ with $T_{3f}=-1/2$ and $b=4$ for $f=u$ with $T_{3f}=+1/2$. The function $A(r)$ is given in \eq{fct:Br}.

\subsubsection{Chargino contribution}
The chargino contributes to the $u$, $d$, and electron EDMs as follows:
\begin{align}
\label{eq:chargf}
    \frac{d^{\tilde\chi^+}_u}{e} &=
    -\frac{\alpha_{\rm em}}{4\pi\sin^2\theta_W}\sum_{k=1}^2\sum_{i=1}^2{\rm Im}[\Gamma_{uik}]\frac{m_{\tilde{\chi}^+_i}}{m^2_{\tilde{d}_k}}\left[Q_{\tilde d}B(m^2_{\tilde{\chi}^+_i}/m^2_{\tilde{d}_k})+(Q_u-Q_{\tilde d})A(m^2_{\tilde{\chi}^+_i}/m^2_{\tilde{d}_k})\right],\\
    \frac{d^{\tilde\chi^+}_d}{e} &=
    -\frac{\alpha_{\rm em}}{4\pi\sin^2\theta_W}\sum_{k=1}^2\sum_{i=1}^2{\rm Im}[\Gamma_{dik}]\frac{m_{\tilde{\chi}^+_i}}{m^2_{\tilde{u}_k}}\left[Q_{\tilde u}B(m^2_{\tilde{\chi}^+_i}/m^2_{\tilde{u}_k})+(Q_d-Q_{\tilde u})A(m^2_{\tilde{\chi}^+_i}/m^2_{\tilde{u}_k})\right],\\
    \frac{d^{\tilde\chi^+}_e}{e} &=
    \frac{\alpha_{\rm em}}{4\pi\sin^2\theta_W}\frac{\kappa_e}{m^2_{\tilde{\nu}_e}}\sum_{i=1}^2 {\rm Im}[U^*_{2i}V_{1i}]m_{\tilde{\chi}^+_i}A(m^2_{\tilde{\chi}^+_i}/m^2_{\tilde{\nu}_e}),
\end{align}
where the function $B(r)$ is given in \eq{fct:Br} and 
\begin{align}
    \Gamma_{uik} &=
    \kappa_u V_{2i}S^*_{d1k}(U^*_{1i}S_{d1k}-\kappa_d U^*_{2i}S_{d2k}),\\
    \Gamma_{dik}&=
    \kappa_d U^*_{2i}S^*_{u1k}(V_{1i}S_{u1k}-\kappa_u V_{2i}S_{u2k}).
\end{align}

\subsection{One loop contributions to the CEDMs}
The light quarks also receive the SUSY contributions to the CEDMs from gluino exchange ($\tilde{d}_q^{\tilde g}$), neutralino exchange ($\tilde{d}_q^{\tilde{\chi}^0}$), and chargino exchange ($\tilde{d}_q^{\tilde{\chi}^+}$), which are respectively given by
\begin{align}
    \tilde{d}_q^{\tilde g} &=
    \frac{\alpha_s}{4\pi}\sum_{k=1}^2{\rm Im}[S^*_{q2k}S_{q1k}]\frac{m_{\tilde g}}{m^2_{\tilde{q}_k}}C(m^2_{\tilde g}/m^2_{\tilde{q}_k}),\\
    \tilde{d}_q^{\tilde{\chi}^0} &=
    \frac{\alpha_{\rm em}}{4\pi\sin^2\theta_W}\sum_{k=1}^2\sum_{i=1}^4{\rm Im}[\eta_{qik}]\frac{m_{\tilde{\chi}^0_i}}{m^2_{\tilde{q}_k}}B(m^2_{\tilde{\chi}^0_i}/m^2_{\tilde{q}_k}),\\
    \tilde{d}_q^{\tilde{\chi}^+} &=
    -\frac{\alpha_{\rm em}}{4\pi\sin^2\theta_W}\sum_{k=1}^2\sum_{i=1}^2{\rm Im}[\Gamma_{qik}]\frac{m_{\tilde{\chi}^+_i}}{m^2_{\tilde{q}_k}}B(m^2_{\tilde{\chi}^+_i}/m^2_{\tilde{q}_k}),
\end{align}
where $C(r)$ is given in \eq{fct:Br}.

\subsection{Total one loop contributions}

For the electron EDM, the SUSY contributions are given by $d_e=d^{\widetilde{\chi}^0}_e + d^{\tilde\chi^+}_e$.
Note that $d_e$ is suppressed by the SUSY breaking scale squared. 
Taking an approximation that all the SUSY particles are much heavier than the electroweak scale, and their masses are typically $\tilde{m}$, $d_e$ is estimated as
\begin{align}
\label{eq:deestimate}
    |d_e| &\sim 10^{-26} ~e\cdot{\rm cm} \times \alpha_{\rm em}\left(
    \frac{10~{\rm TeV}}{\tilde m}
    \right)^2|\tan\beta \sin\theta_\mu - \sin\theta_A|,
\end{align}
which is a more simplified version of the formula in Eq.~28 of \cite{Falk:1999tm} and aims to emphasize the possible cancellation between the phases $\theta_\mu$ and $\theta_A$. 
In a similar manner, the quark EDM is given by $d_q =d_q^{\tilde g} + d^{\widetilde{\chi}^0}_e + d^{\tilde\chi^+}_e$ and is estimated as 
\begin{align}
    |d_q| \sim 10^{-26}~e\cdot{\rm cm}\times \alpha_s\left(
    \frac{10~{\rm TeV}}{\tilde m}
    \right)^2
    \times
    \begin{cases}
        |\cot\beta\sin\theta_\mu - \sin\theta_A| & (q=u)\\
        |\tan\beta\sin\theta_\mu - \sin\theta_A| & (q=d)
    \end{cases}
    ,
\end{align}
where the contributions from the gluino exchange dominate over those from the neutralino and chargino exchange.
The CEDM is given by $\tilde d_q=\tilde{d}^{\tilde g}_q+\tilde{d}^{\tilde{\chi}^0}_q+\tilde{d}^{\tilde{\chi}^+}_q$ and is estimated as
\begin{align}
    |\tilde d_q| \sim 10^{-25}~{\rm cm}\times \alpha_s\left(
    \frac{10~{\rm TeV}}{\tilde m}
    \right)^2
    \times
    \begin{cases}
        |\cot\beta\sin\theta_\mu - \sin\theta_A| & (q=u)\\
        |\tan\beta\sin\theta_\mu - \sin\theta_A| & (q=d)
    \end{cases}.
\end{align}
With these contributions, once the SUSY spectrum is obtained as explained in the next section, we may obtain the equivalent electron EDM through Eq.~(\ref{eq:deequiv}) and the neutron EDM through Eq. (\ref{eq:dnnum}). It is important note that from the structure of $d^{\tilde\chi^0}_u$ in \eq{eq:neutf} and $d^{\tilde\chi^+}_u$ in \eq{eq:chargf}, together with the simplified expression in \eq{eq:deestimate}, we see that chargino and neutralino contributions have opposite signs. However, even if a total cancellation was possible at 1-loop level, the 2-loop contributions become important, particularly from $d_f^{\, \gamma A^0 \tilde{f}}$ as is well known \cite{Pilaftsis:1999td}. In Appendix \ref{sec:2lsusy} we give a complete list of 2-loop contributions and compare them with the 1-loop contributions in Figs.~(\ref{fig:tanb_vs_deEDM}-\ref{fig:M12_vs_deEDM}).

%%%%%%%%%%%%%%%%%%
\section{Results}
\label{results}
%%%%%%%%%%%%%%%%%%

We will apply our calculations of EDMs to the CMSSM \cite{cmssm}.  These models are defined by four continuous parameters. 
These are: a universal gaugino mass, $m_{1/2}$, a universal scalar mass, $m_0$, a universal trilinear term, $A_0$, and the ratio of the two Higgs vacuum expectation values, $\tan \beta$.
In the CMSSM, universality is often assumed to occur at the same renormalization scale for which gauge coupling unification occurs, namely when the two electroweak gauge couplings are equal, $g_1(Q) = g_2(Q)$. The scalar masses, gaugino masses, and $A$-terms are run down from the Grand Unified (GUT) scale to the electroweak scale and minimization of the Higgs potential fixes 
the $\mu$-term as well as the bi-linear term, $B$, which can be related to the heavy pseudo-scalar Higgs boson mass. 
In the minimal CMSSM, the sign of the $\mu$-parameter is left free. Of course in the present context, the arguments of $\mu$ and $A_0$ are taken to be two additional free parameters which will be used to compute the electric dipole moments under consideration.

\begin{figure}[ht!]
    \centering
    \includegraphics[width=7cm]{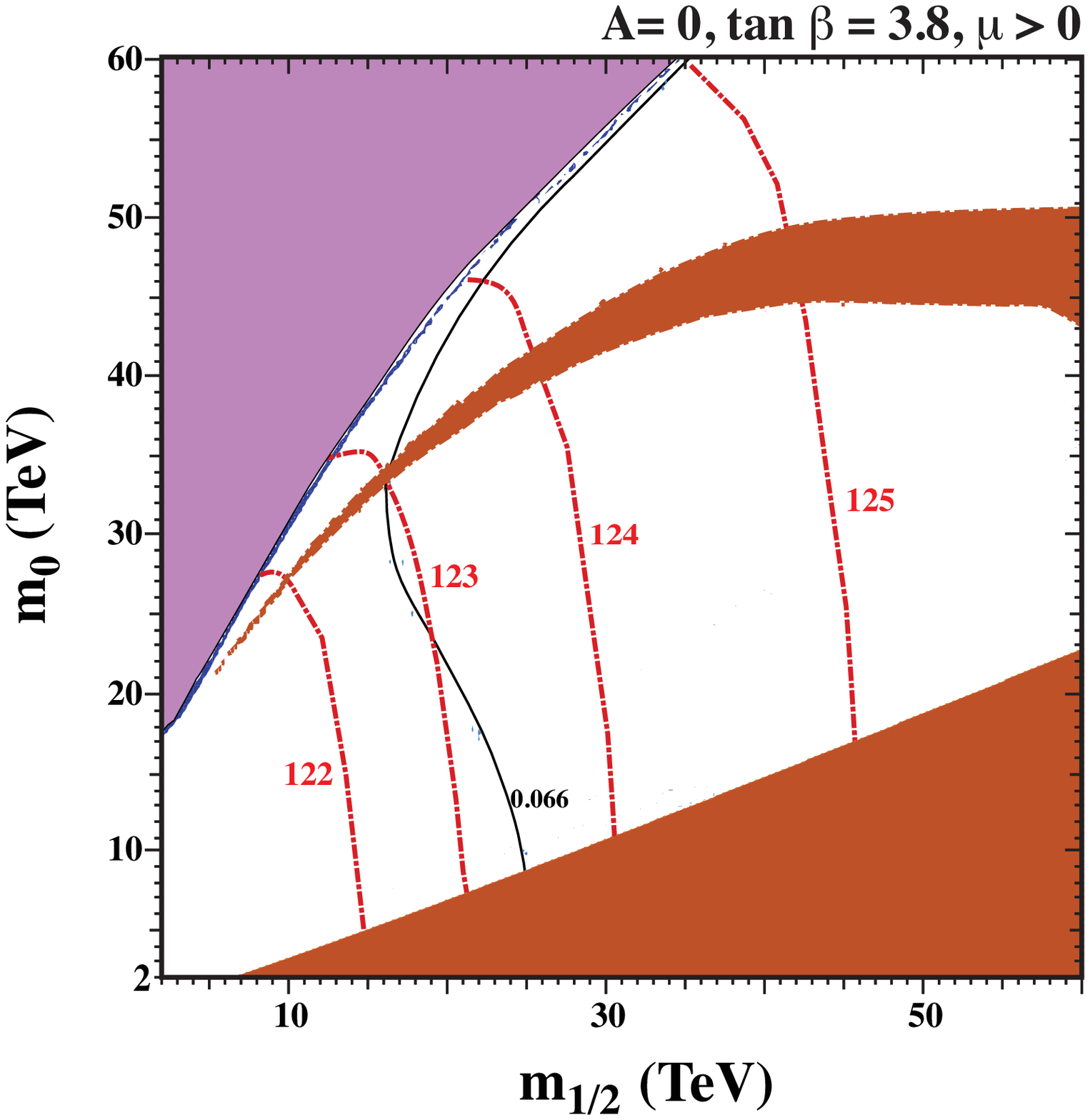}
     \includegraphics[width=7cm]{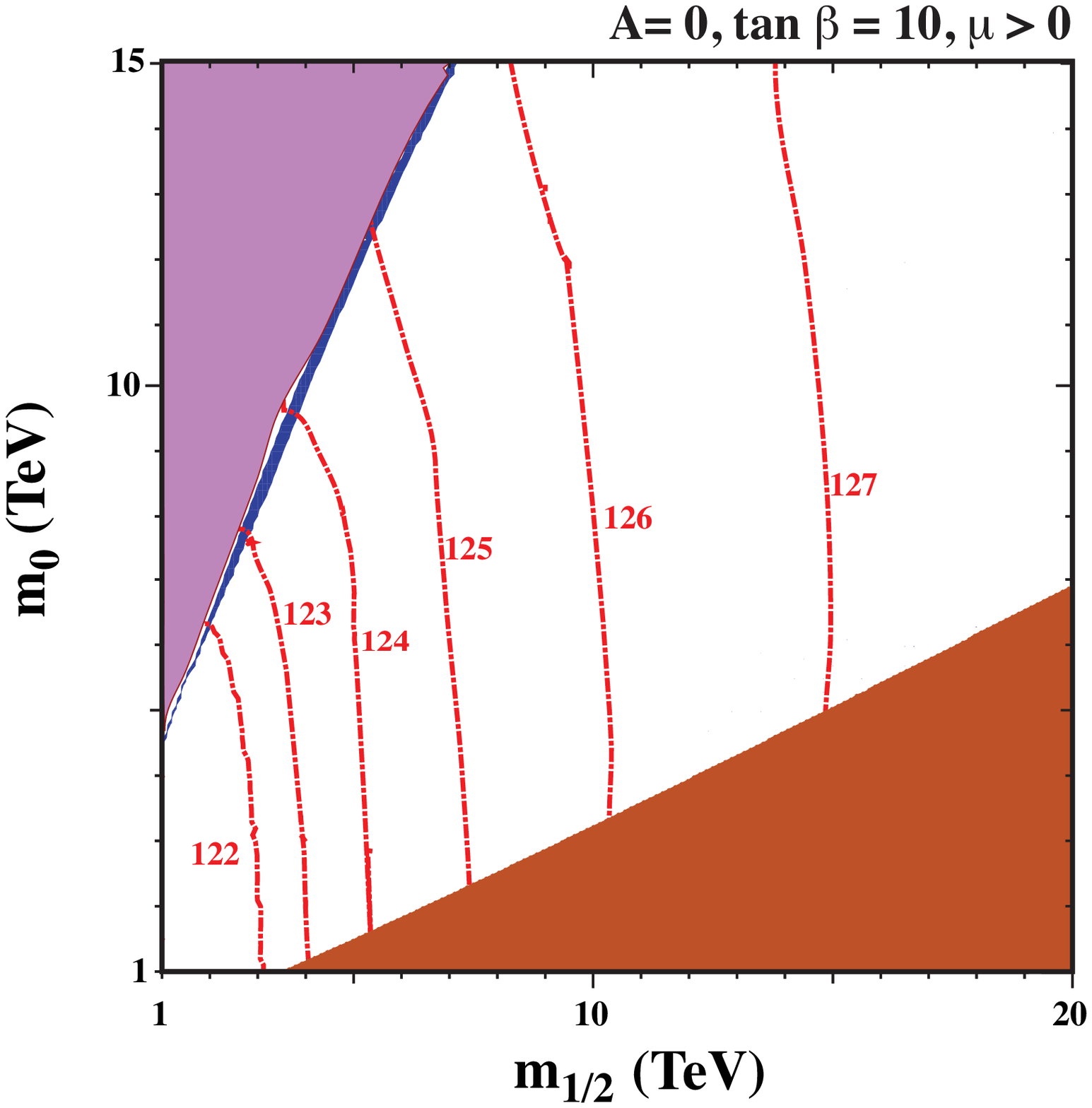} \\
     \includegraphics[width=7cm]{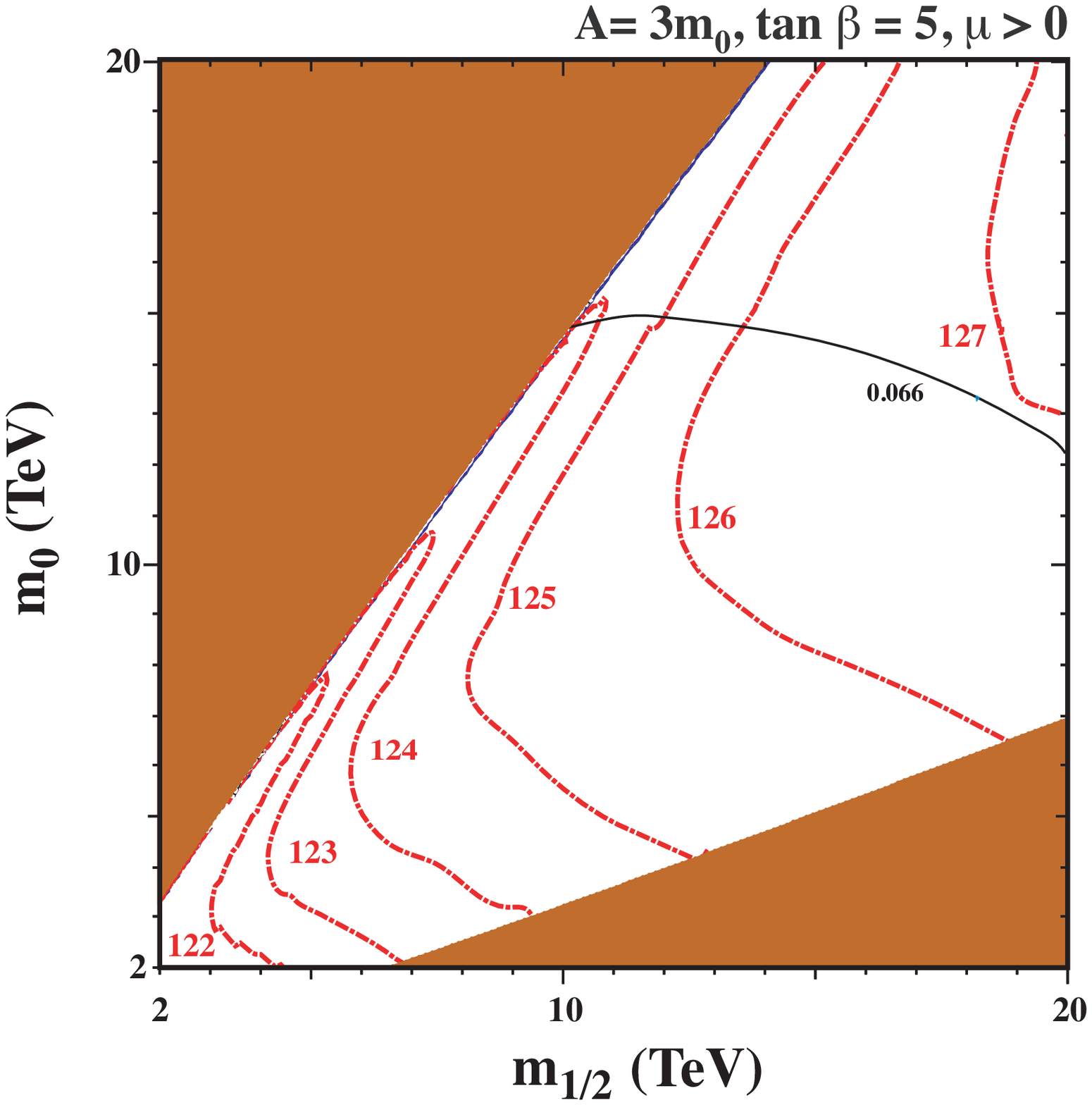} 
      \includegraphics[width=7cm]{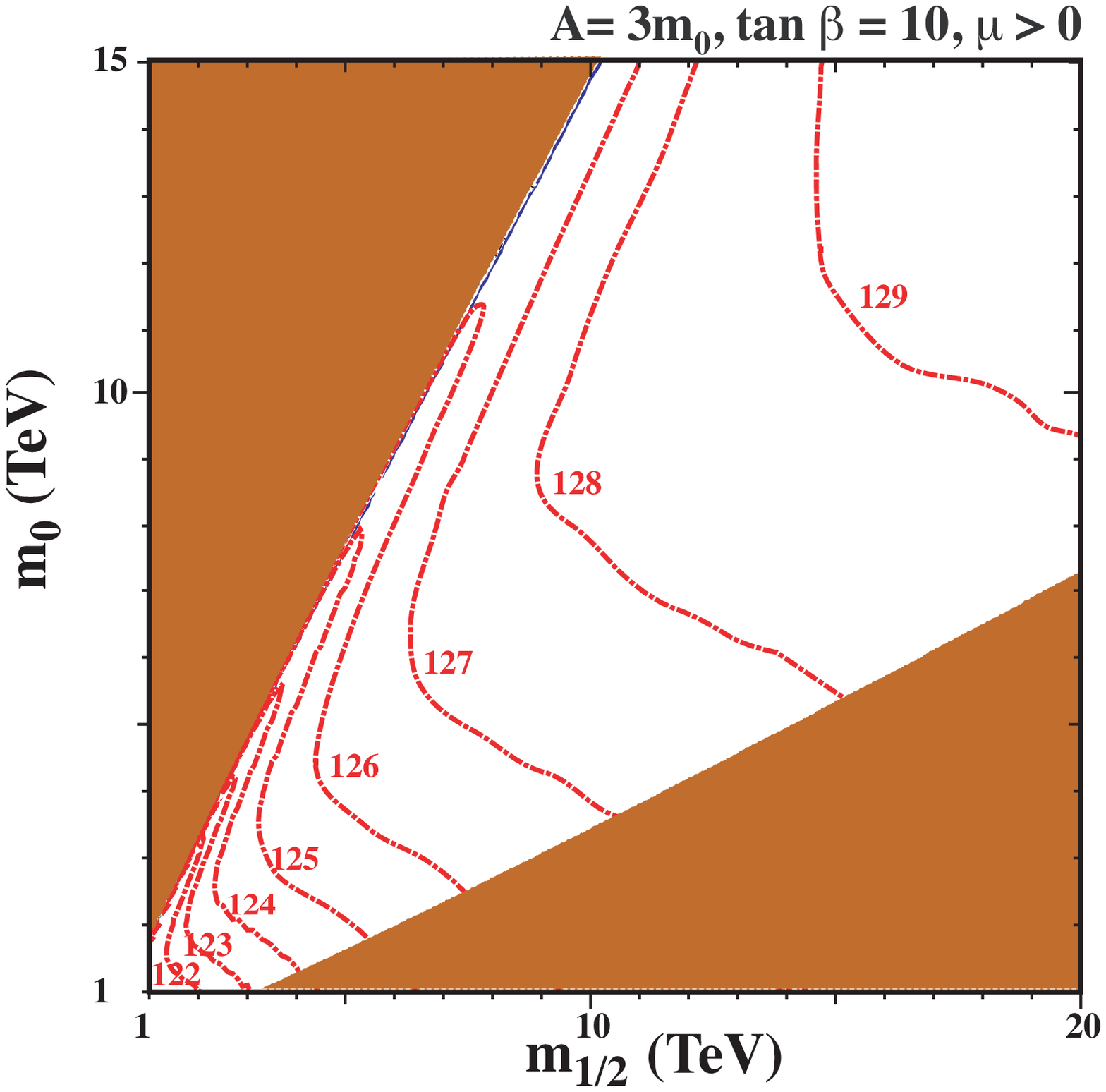}
    \caption{The $(m_{1/2}, m_0)$ plane for $A_0 = 0$ and $\tan \beta = 3.8$ (case A - upper left); $A_0 = 0$ and $\tan \beta = 10$ (case B - upper right); $A_0 = 3 m_0$ and $\tan \beta = 5$ (case C - lower left); $A_0 = 3 m_0$ and $\tan \beta = 10$ (case D - lower right) all with $\mu > 0$ in the CMSSM with GUT-scale universality. Here all CP-violating phases are set to 0.  The red dot-dashed contours show the lightest Higgs mass as calculated using {\tt FeynHiggs}~\cite{FH}. In the dark red shaded regions, the LSP is either lighter stau, stop or chargino as described in the text. In the blue
strip along either the region with no electroweak symmetry breaking (shaded pink in the upper two panels) or the stop LSP region, the LSP has an enlarged relic density range, $0.02 < \Omega_\chi h^2 < 0.6$, to enhance its visibility.  The black lines seen in both of the left panels show the constraint from the lifetime of $p \to K^+ \bar{\nu}$.}
    \label{fig:cmssm}
\end{figure}

In Fig.~\ref{fig:cmssm},
we show four representative $m_{1/2},m_0$ planes for fixed $A_0$ and $\tan \beta$, all with $\mu > 0$. In these panels, the 
phases are all set to zero. We refer to these possibilities as:
\begin{itemize}
\item A (top left): $A_0 = 0$ and $\tan\beta=3.8$,
\item B (top right): $A_0 = 0$ and $\tan\beta=10$,
\item C (bottom left): $A_0 = 3 m_0$ and $\tan\beta=5$,
\item D (bottom right): $A_0 = 3 m_0$ and $\tan\beta=10$.
\end{itemize}
These planes are chosen so as to sample the parameter space in regions which contain a Higgs boson with mass $m_H \approx 125$ GeV \cite{ATLAS:2012yve,CMS:2012qbp} as well as containing a dark matter candidate with a relic density $\Omega_\chi h^2 \approx 0.12$ \cite{Planck}. 
In the pink shaded region in the upper left corner of the upper two panels (cases A and B), radiative electroweak symmetry breaking does not occur as the $\mu^2$ runs to negative values. 
In the dark red shaded regions seen in all of the panels, the lightest supersymmetric particle (LSP) is charged and therefore excluded. In the regions found in the lower right corners, the LSP is the lighter stau, whereas in the upper left (for cases C and D), the LSP is the lighter stop. For case A, there is a region in the middle of the panel where the LSP is the lighter chargino. These regions are all excluded. 

The panels also show contours  (red dot-dashed) giving the value of the mass of the light Higgs scalar, $m_H$ as labeled (in GeV). These are computed using {\tt FeynHiggs 2.18.1} \cite{FH}. The thin blue region shows the parameter values where the relic density of the LSP is in agreement with that determined by Planck \cite{Planck}. To make the strips more visible on the scale of these plots, we shade regions with a relic density $\Omega_\chi h^2 = (0.06 - 0.2)$. These strips run along the border where radiative electroweak symmetry breaks down (when $A_0 = 0$) known as the focus point region \cite{fp} or the border of the stop LSP region \cite{stopco} when $A_0/m_0$ is relatively large. Finally in cases A and C, we also show the limit from proton lifetime ($\tau(p \to K^+ \nu) = 6.6 \times 10^{33}$~y~\cite{Abe:2014mwa}) by a black solid curve. Points to the left (or below) these lines have shorter lifetimes, assuming minimal SU(5).  For the other cases shown, because of the larger values of $\tan \beta$, the proton lifetime is too short if one assumes minimal SU(5). For more on these types of planes, see \cite{recent}. 

\begin{figure}[ht!]
    \centering
    \includegraphics[width=7cm]{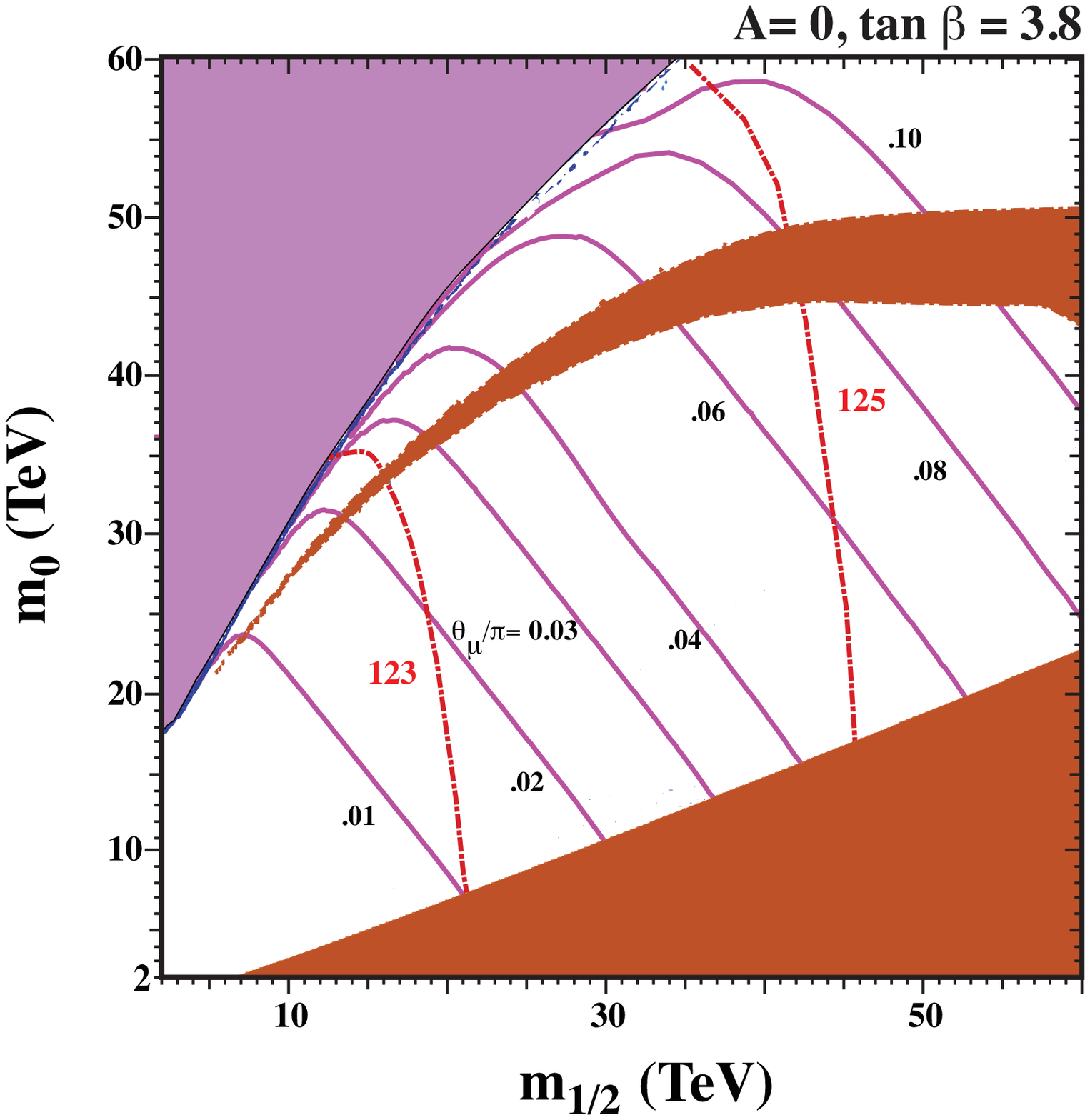}
     \includegraphics[width=7cm]{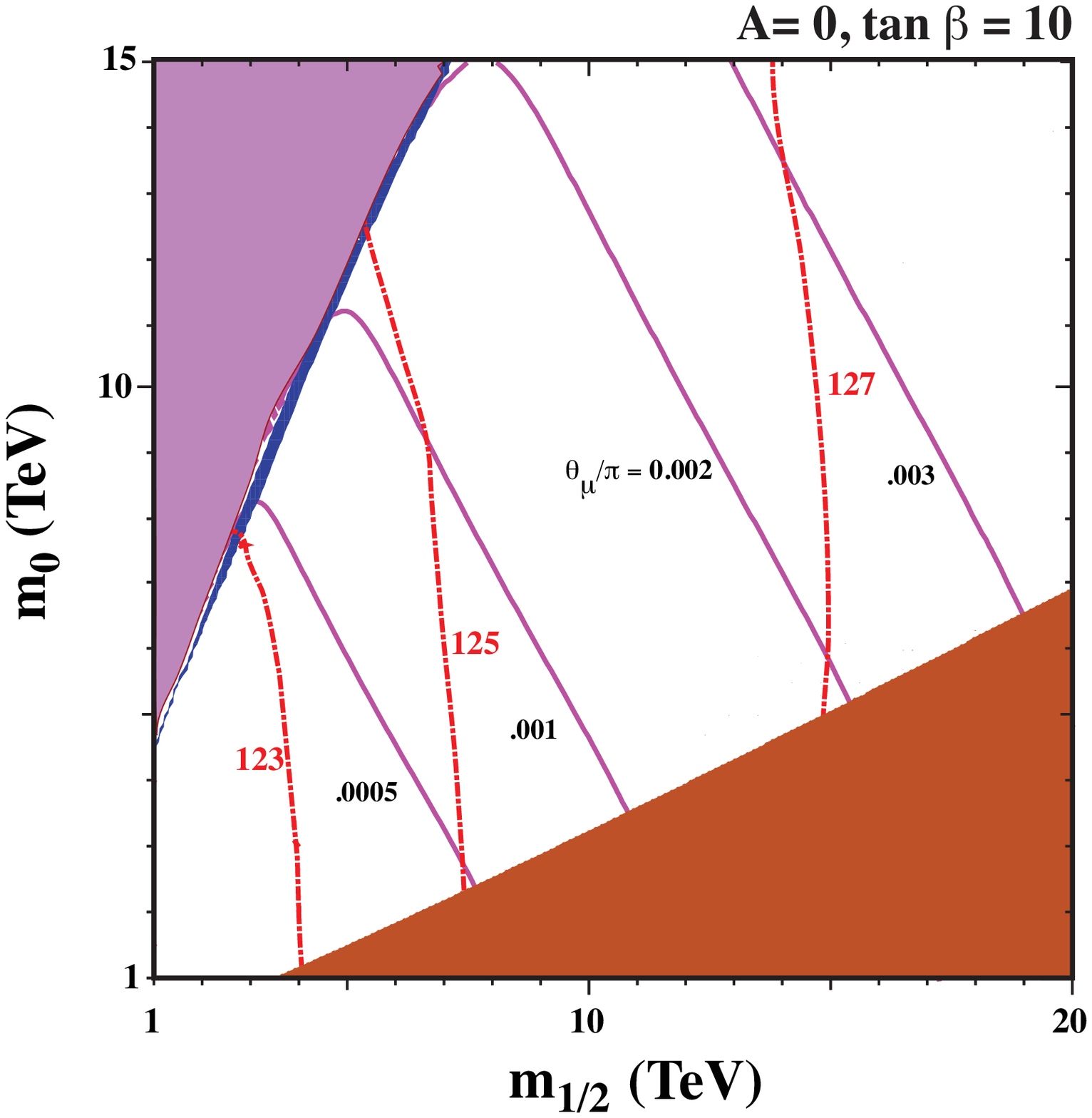} \\
     \includegraphics[width=7cm]{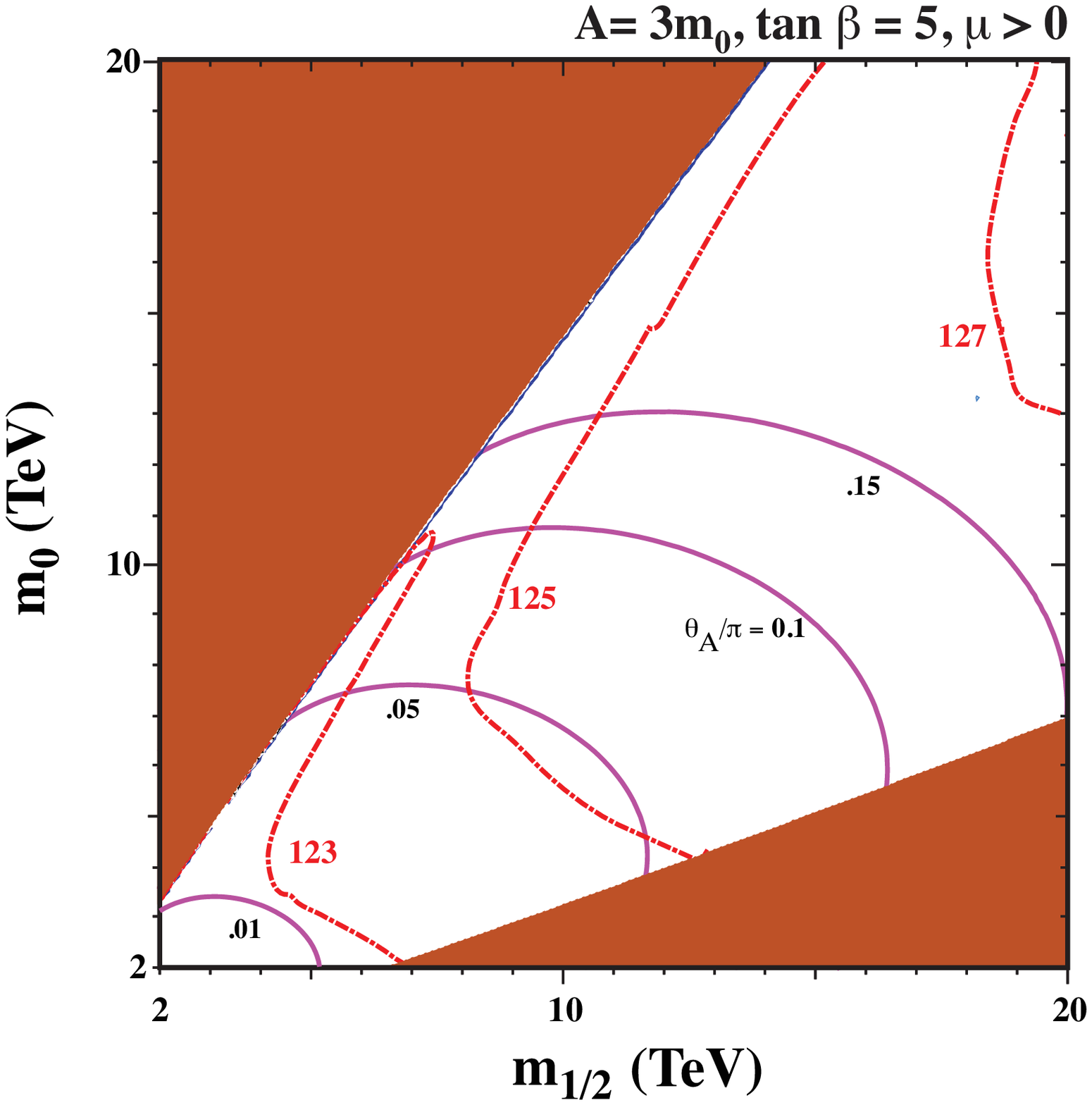}       \includegraphics[width=7cm]{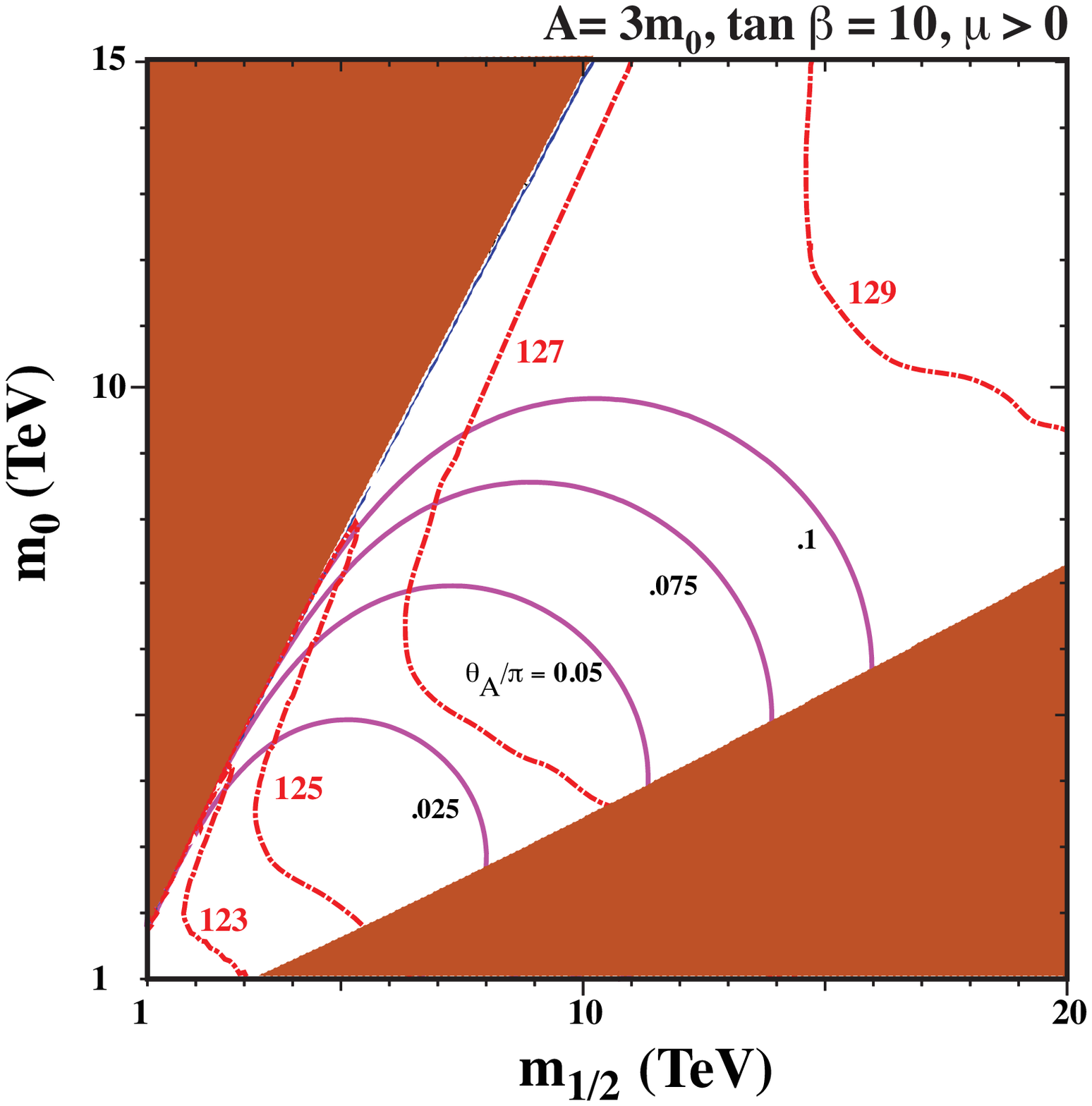}
    \caption{Limits from the electron EDM for several values of $\theta_\mu$ (upper panels) and several values of $\theta_A$ (lower panels) for the same parameter planes shown in Fig.~\ref{fig:cmssm}. Regions of parameter space below and/or to the left of each the solid blue curves is excluded for that value of $\theta_\mu$ or $\theta_A$. 
    }
    \label{fig:eEDM}
\end{figure}

\begin{figure}[ht!]
    \centering
    \includegraphics[width=7cm]{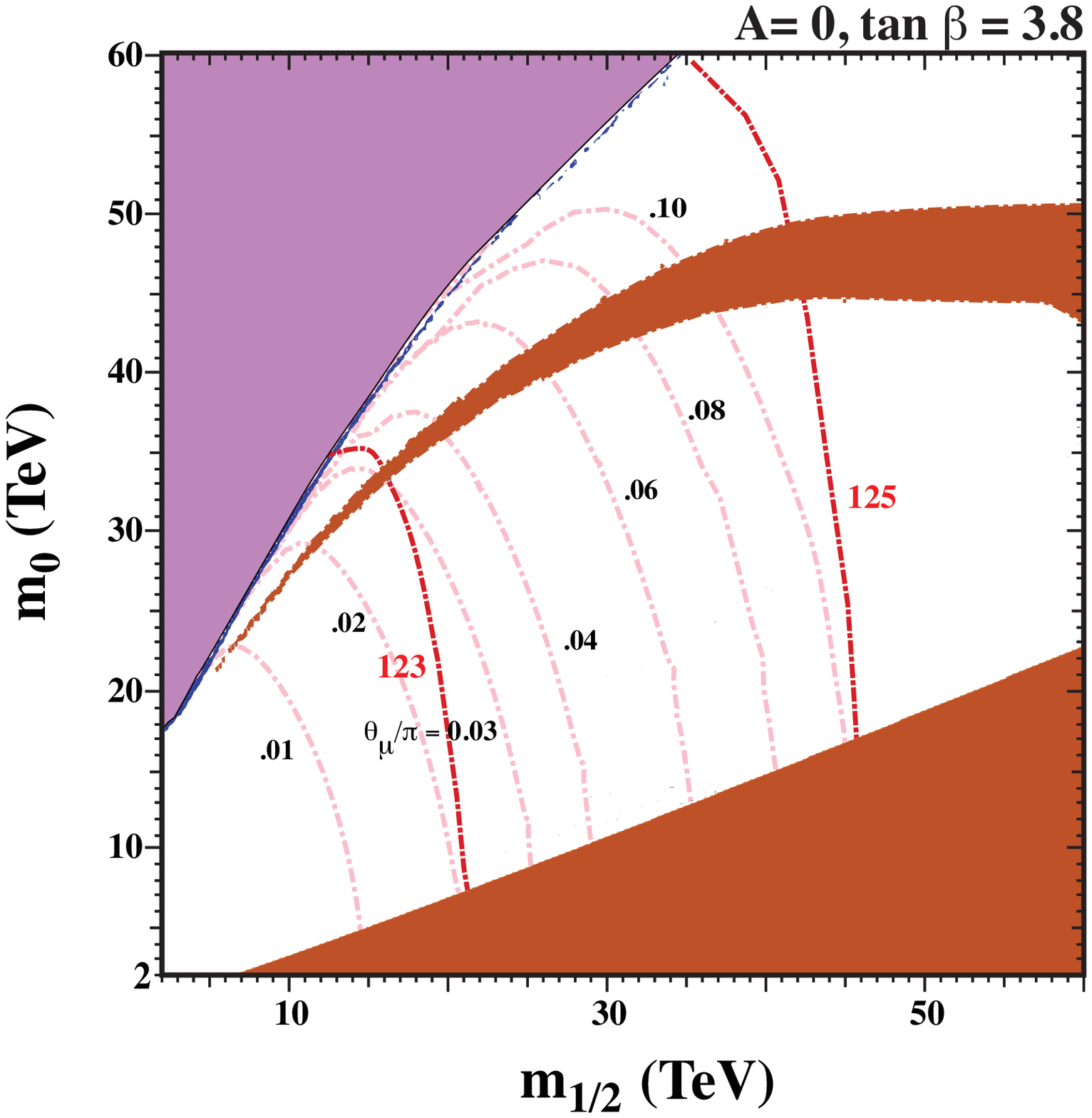}
     \includegraphics[width=7cm]{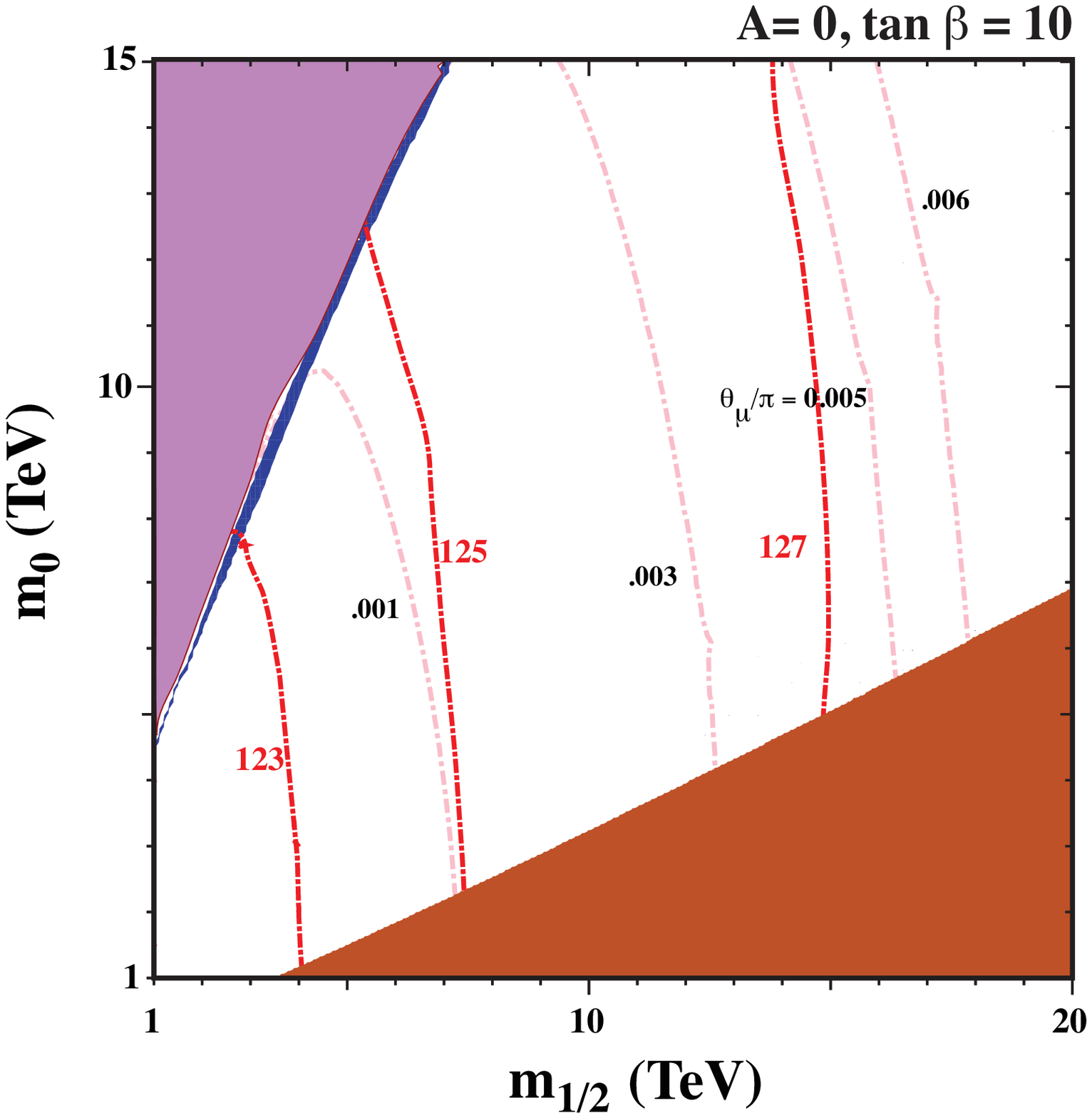} \\
     \includegraphics[width=7cm]{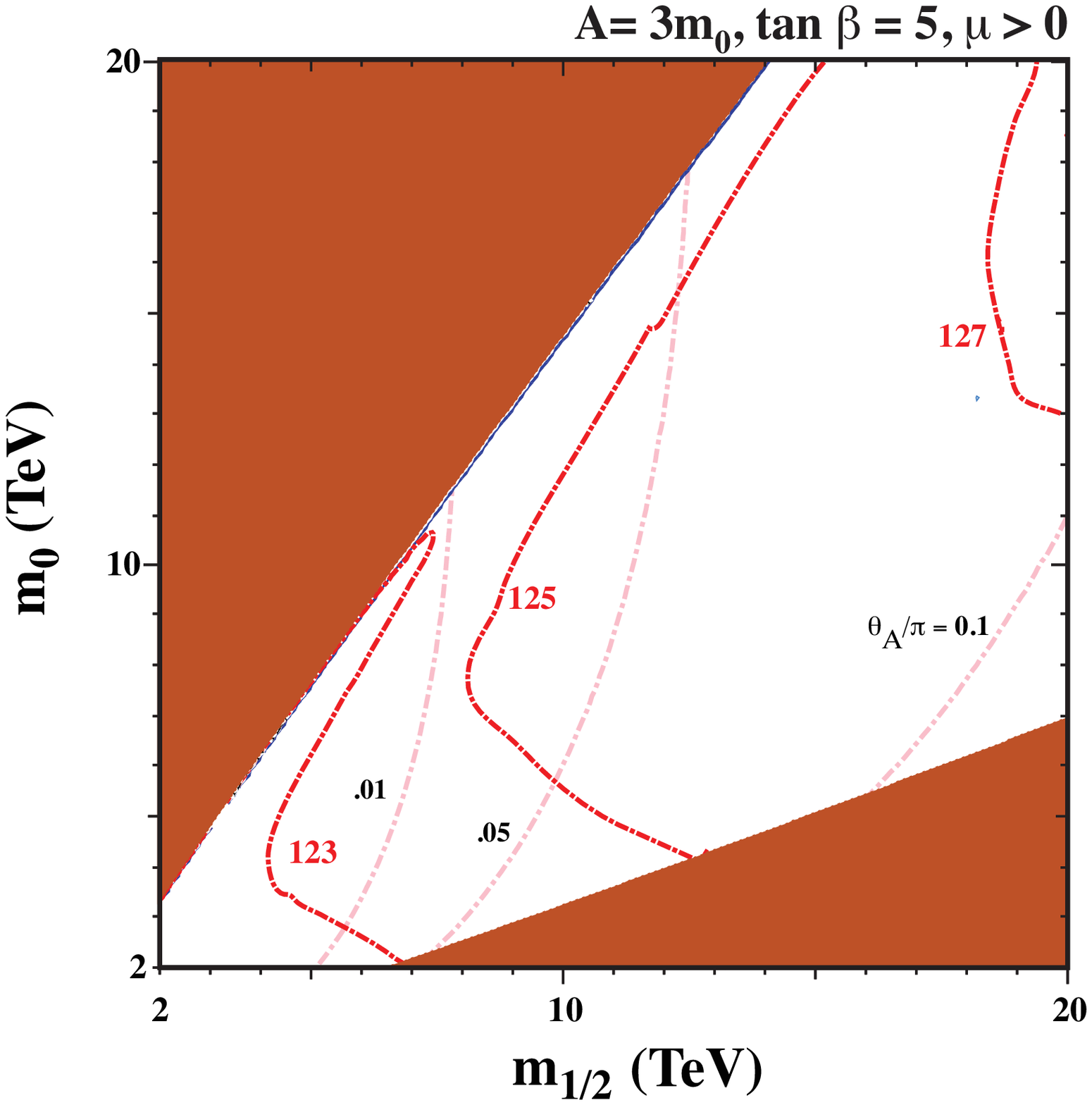} 
     \includegraphics[width=7cm]{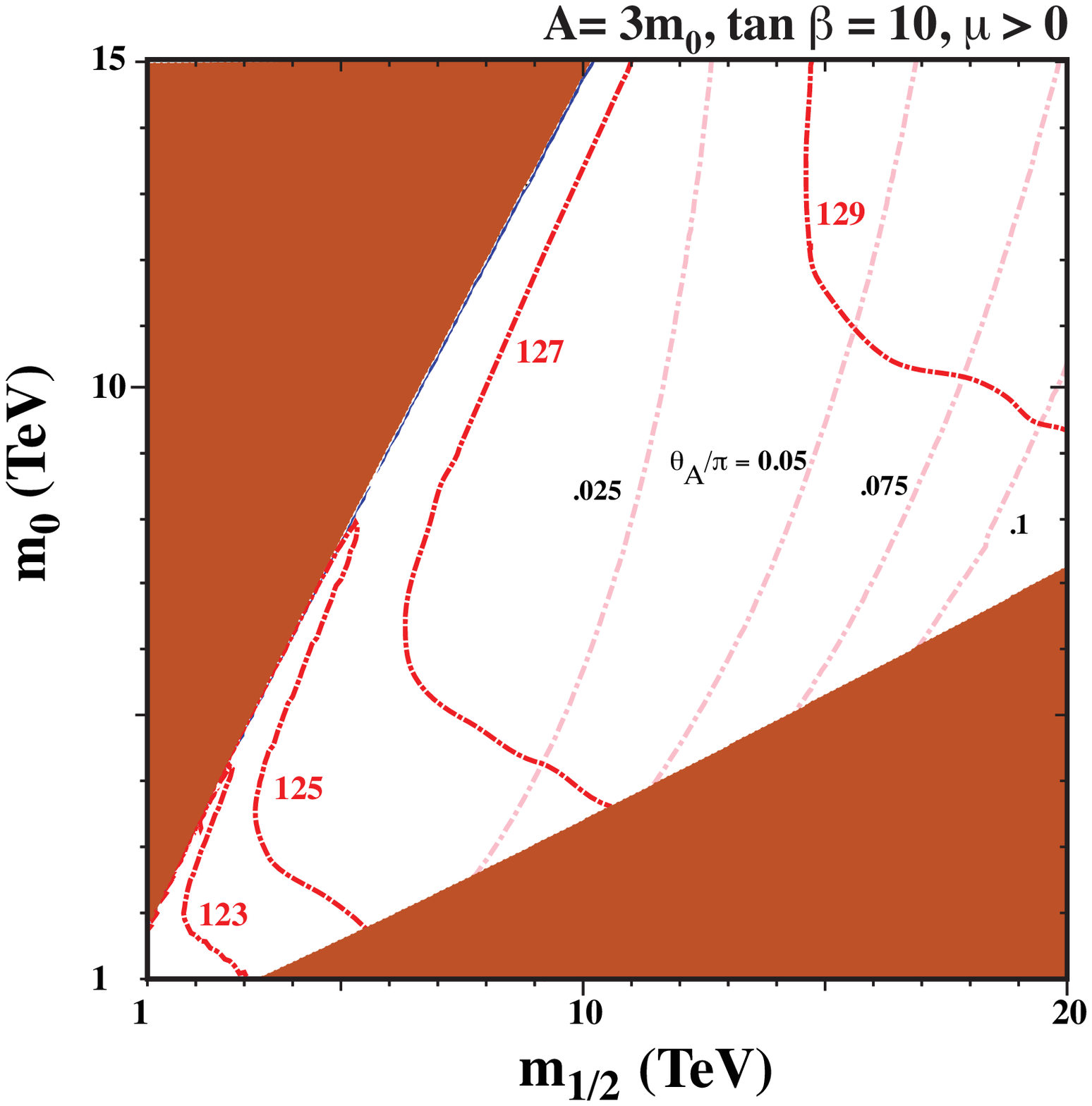}
    \caption{As in Fig.~\ref{fig:eEDM} using the neutron EDM 
    assuming that the current neutron EDM limit is multiplied by $10^{-3}$. 
    Limits on $\theta_\mu$ (upper panels) and $\theta_A$ (lower panels) are shown by pink dot dashed curves. }
    \label{fig:nEDM}
\end{figure}

\begin{figure}[ht!]
    \centering
    \includegraphics[width=7cm]{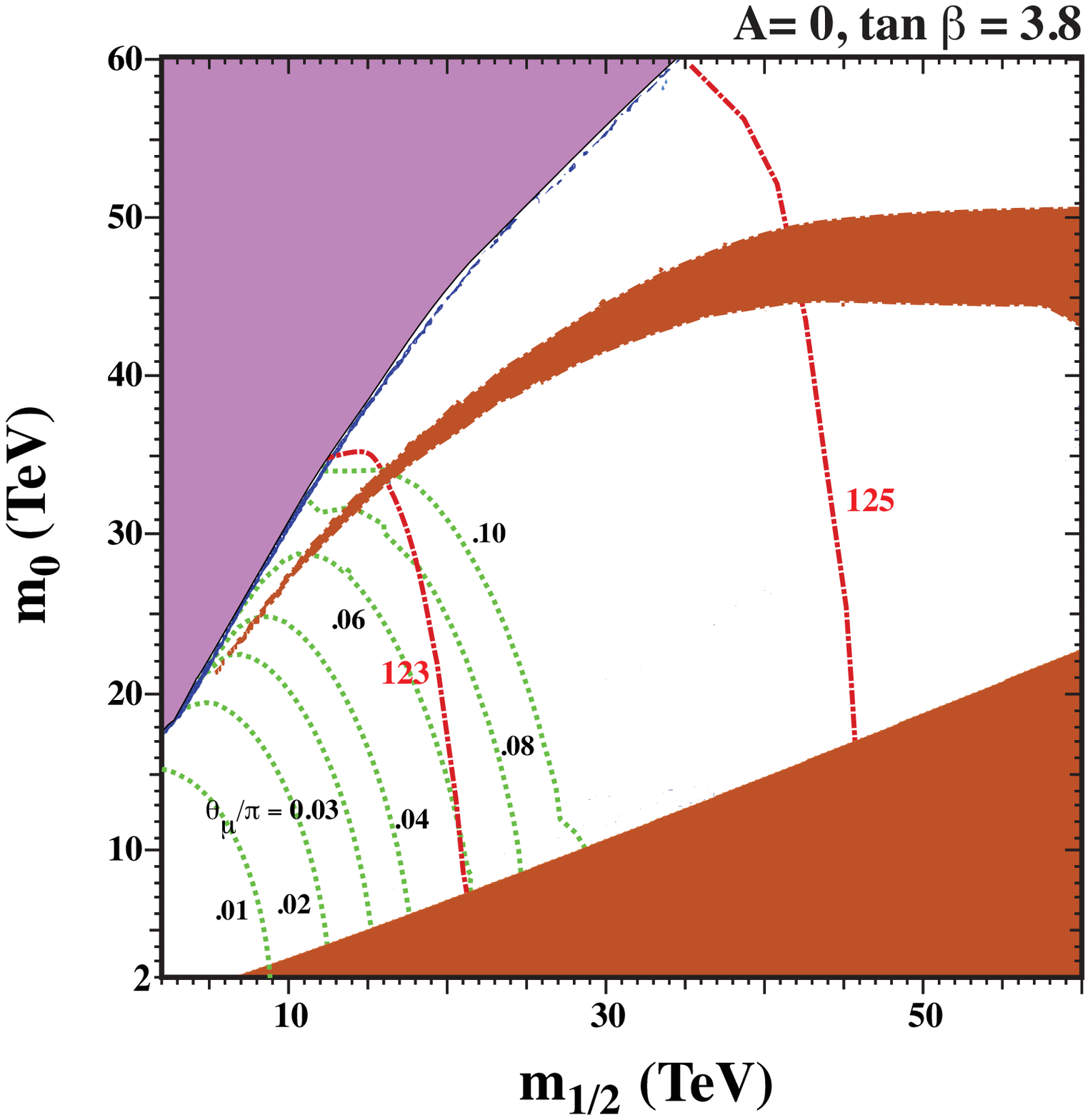}
     \includegraphics[width=7cm]{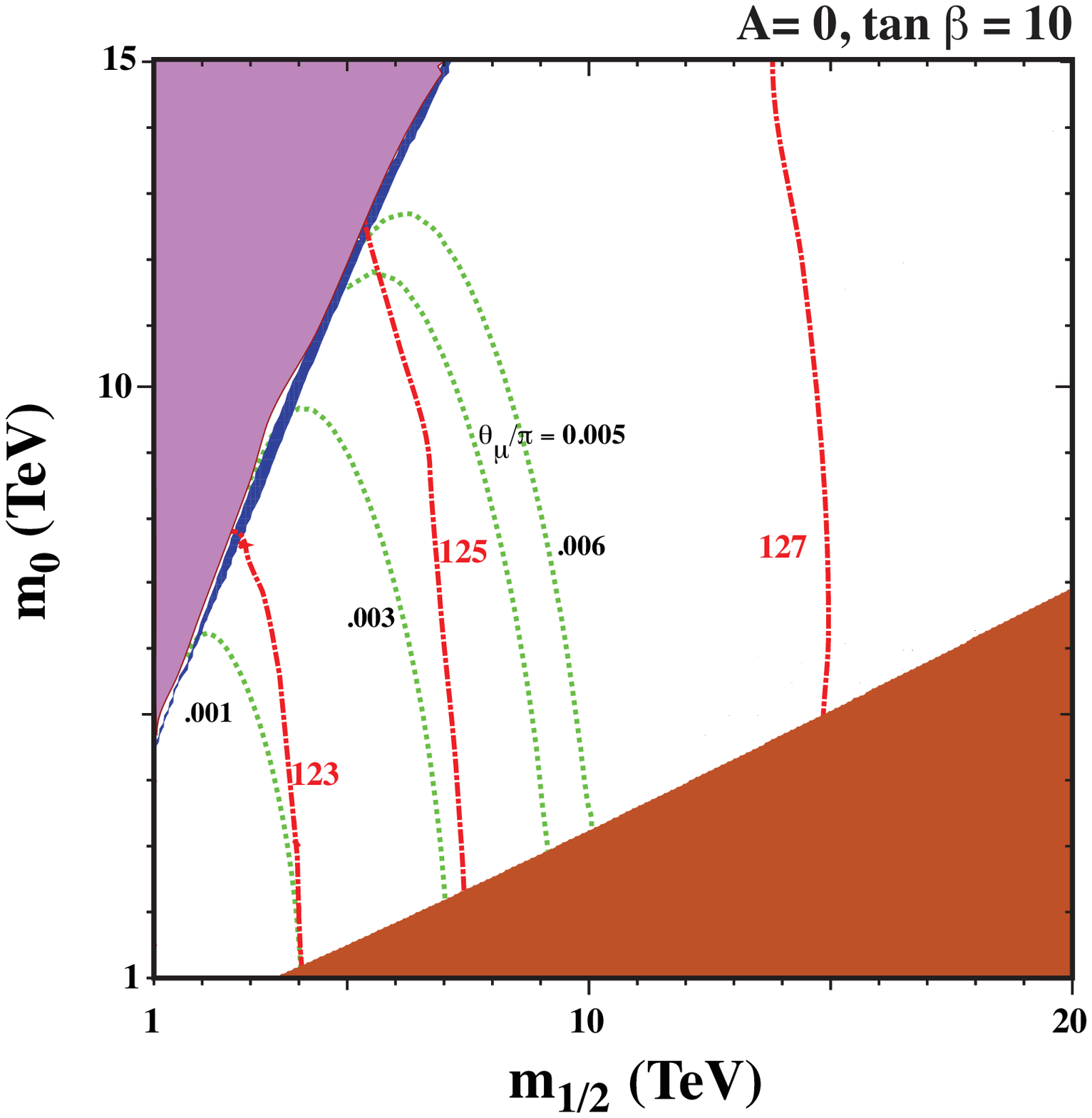} \\
    \includegraphics[width=7cm]{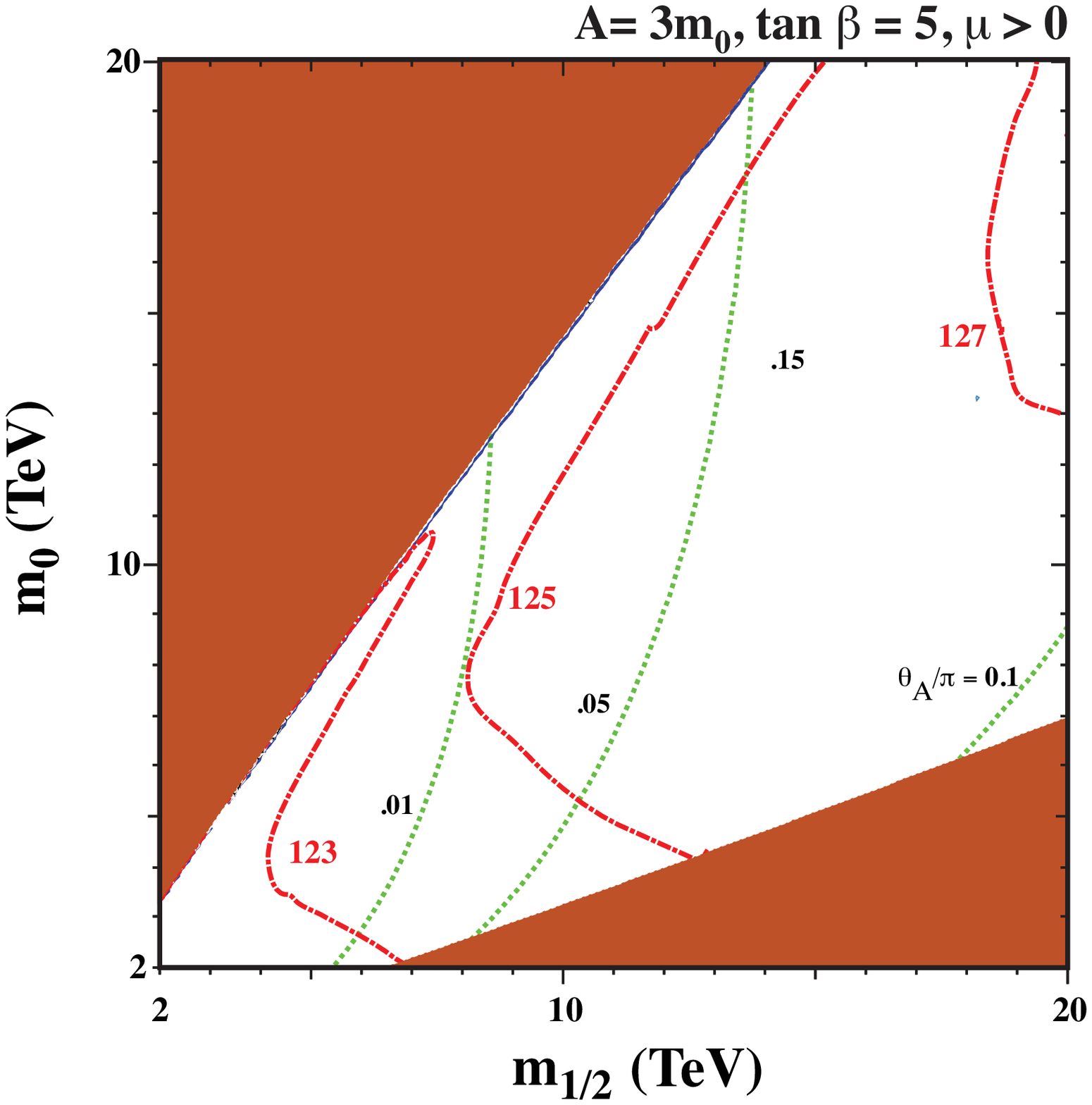}       \includegraphics[width=7cm]{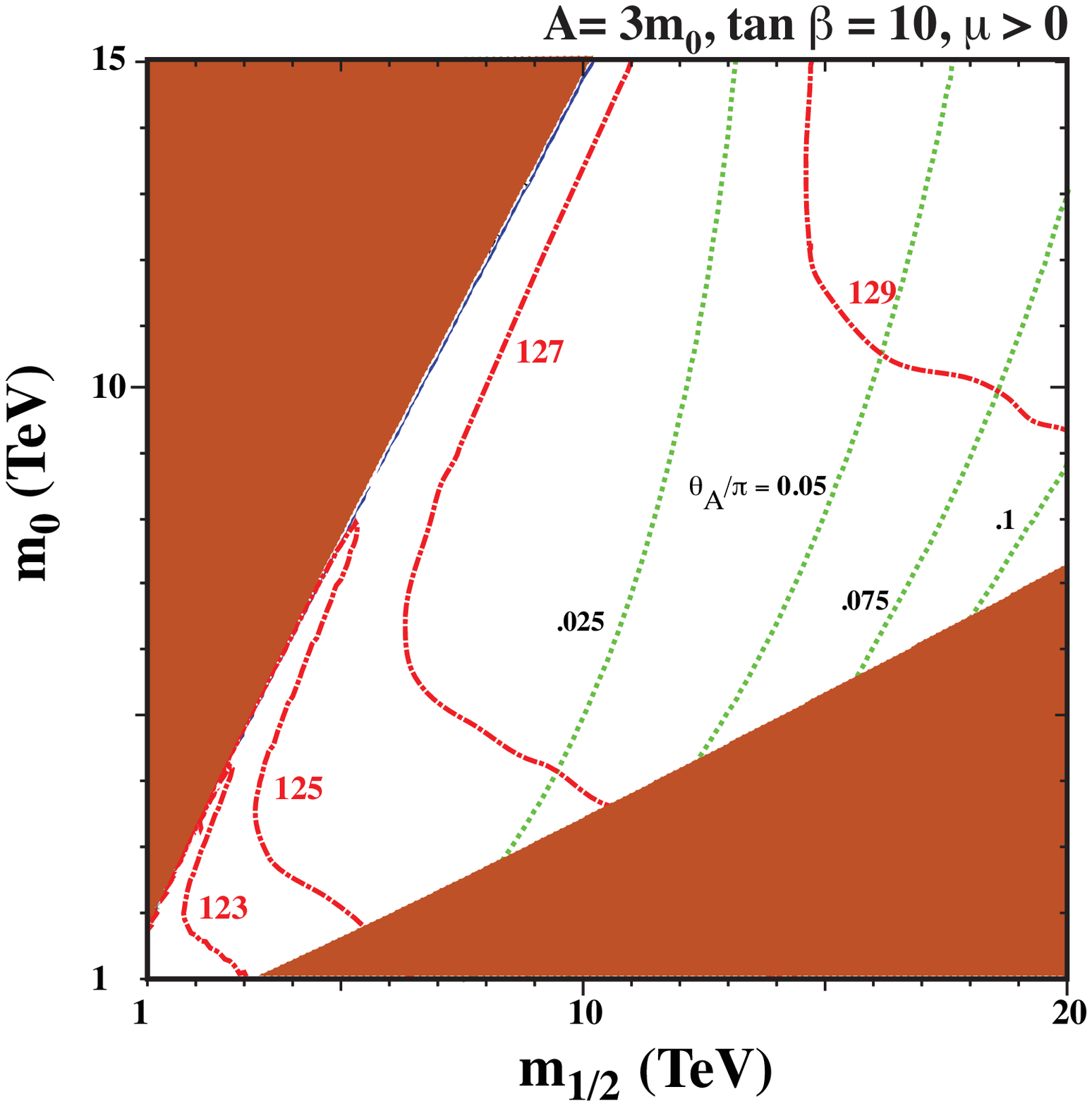}
    \caption{As in Fig.~\ref{fig:eEDM} using the proton EDM 
    assuming $d_p <  10^{-29}~e \cdot \mathrm{cm}$. 
    Limits on $\theta_\mu$ (upper panels) and $\theta_A$ (lower panels) are shown by green dotted curves.}
    \label{fig:pEDM}
\end{figure}

Using these examples, we can derive the constraints from the EDMs when the phases $\theta_\mu$ and $\theta_A$ are non-zero. In Figs.~\ref{fig:eEDM}, \ref{fig:nEDM}, and \ref{fig:pEDM} we show the constraints imposed by experimental limits for a fixed phase. In Fig.~\ref{fig:eEDM}, we apply the constraints from the electron EDM to each of the parameter planes for cases A-D. 
For cases A and B, $\theta_A = 0$ and contours for fixed $\theta_\mu/\pi$ (as labeled) are shown by the solid magenta lines. The area below and/or to the left of the lines are excluded, as the electron EDM surpasses the experimental limit, taken here to be $d_e < 4.1 \times 10^{-30}~e \cdot \mathrm{cm}$ \cite{Roussy:2022cmp}. For example, in case B 
when $\theta_\mu/\pi = 0.002$, much of the parameter space is excluded, including parameter values which lead to a Higgs mass of 125 GeV. To be compatible with both the relic density and the Higgs mass in this case, the phase of $\mu$ must be smaller than about $0.0012 \pi$. For case A, because of the relatively low value of $\tan \beta$, larger phases are allowed, however convergence of the RGEs prevents us from adopting phases much larger than about $0.1\pi$. 

In the lower panels of Fig.~\ref{fig:eEDM}, we take $\theta_\mu = 0$ and we see that phases of $A_0$ of order $10^{-1}\pi$ are allowed. For example, in case D, values of $\theta_A > 0.1\pi$ would exclude parameters which yield $m_H = 125$ GeV with the correct relic density. Thus despite the very large mass scales depicted ($\mathcal{O}(10)$ TeV), the improvements in the experimental limit on $d_e$ still allow one to set meaningful constraints on the CP-violating phases in the CMSSM. 

In Figs.~\ref{fig:nEDM} and \ref{fig:pEDM}, we show similar constraints from the neutron and proton EDMs respectively {\em assuming} a factor of $10^3$ improvement in current limit for the neutron EDM, i.e. $d_n < 1.8 \times 10^{-29}~e \cdot \mathrm{cm}$ and a prospective limit on $d_p$ of order  $\mathcal{O}(10^{-29})~e \cdot \mathrm{cm}$ \cite{Alexander:2022rmq}, or approximately a factor of $2 \times 10^4$ improvement for the proton EDM. 
Currently limits from these EDMs would not permit us to set any constraints on the parameter space, and only with an improvement by a factor of about $10^3$ will see comparable constraints to that imposed by $d_e$. 

\begin{figure}[ht!]
    \centering
    \includegraphics[width=7cm]{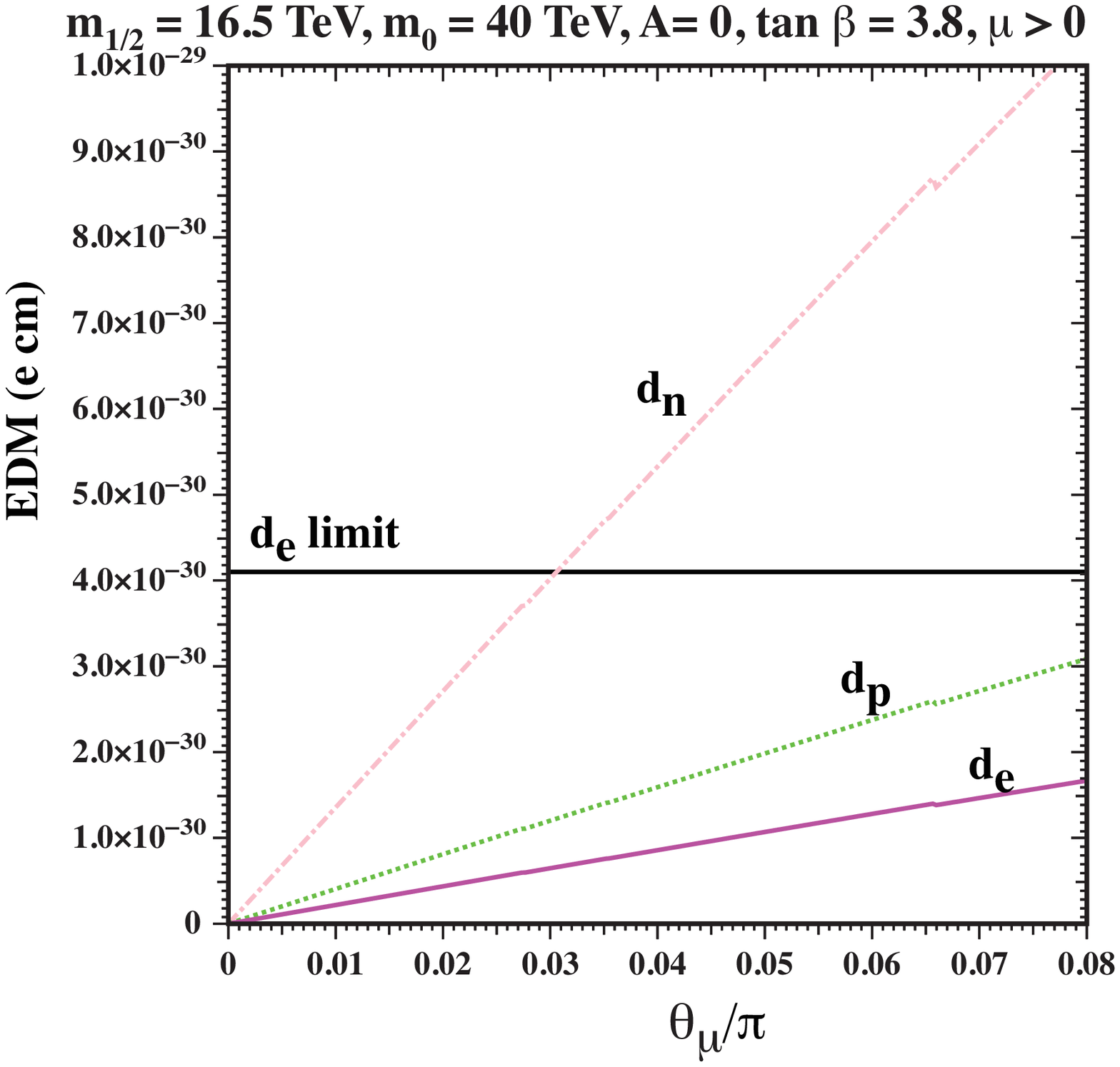}
     \includegraphics[width=7cm]{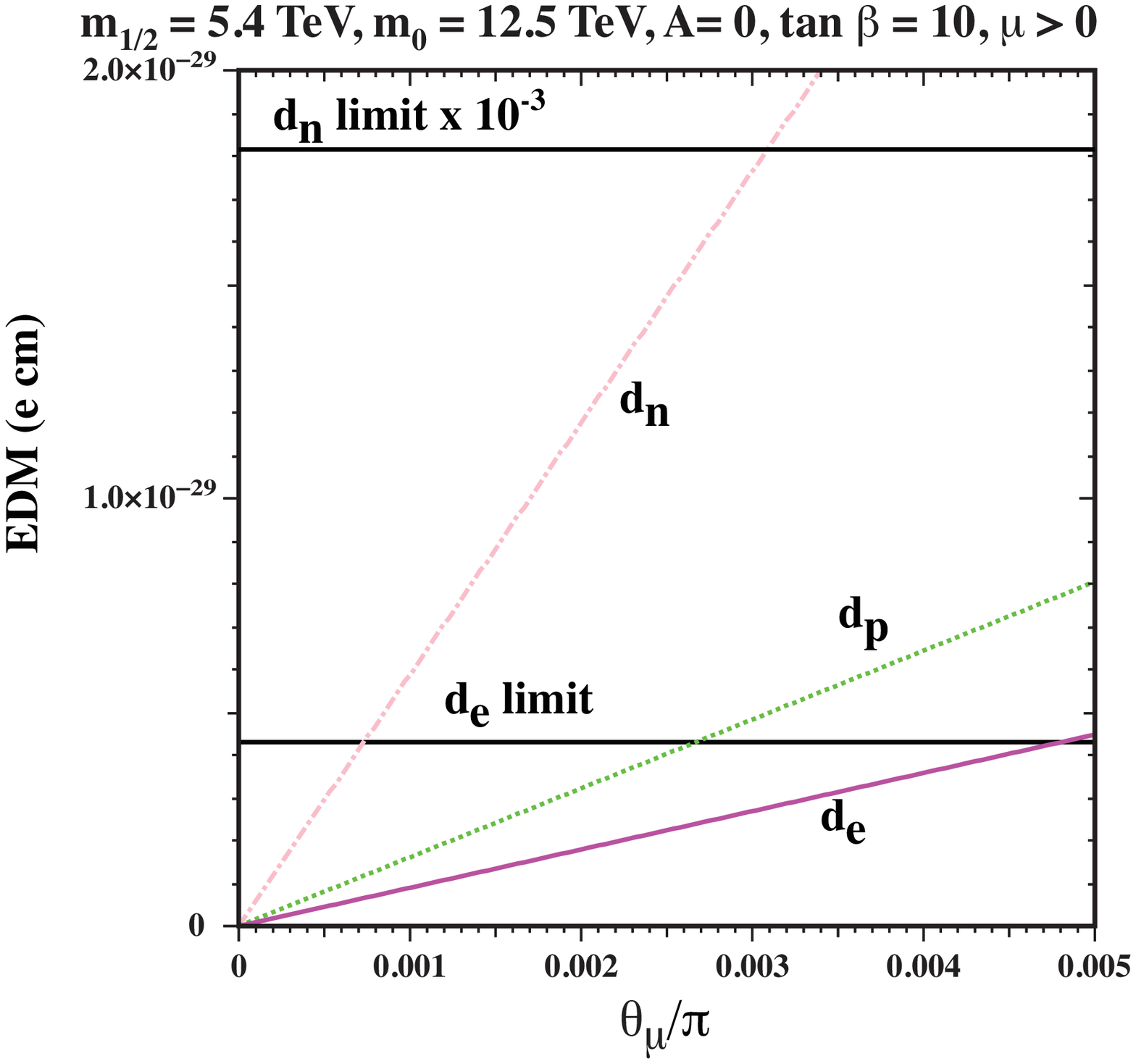} %\\
    \caption{$e$, $p$, and $n$ EDMs as a function of $\theta_\mu$ for fixed $m_{1/2}, m_0, A_0$ and $\tan \beta$ as labelled. }
    \label{fig:thmu}
\end{figure}

In Fig.~\ref{fig:thmu}, we show the dependence of the three EDMs as a function of $\theta_\mu$ for two benchmark points associated with case A (left) and case B (right). These were chosen so that when $\theta_\mu = 0$, the relic density and Higgs mass take acceptable values.
For Case A, $m_{1/2} = 16.5$ TeV, $m_0 = 40$ TeV, $A_0 = 0$, and $\tan \beta = 3.8$. In this case (because of the relatively  low value of $\tan \beta$), the Higgs mass is found to be 123.5 GeV, which
given the uncertainty in the calculation is within the margin of error. All phases up to $0.08\ \pi$  are acceptable, and an improvement in both the experimental limits on $d_e$ and $d_n$ are necessary to constrain the parameter space. Note that for these inputs the running of the RGE's becomes unreliable at higher phases. The horizontal lines correspond to the experimental limits  though the limit for $d_n$ is assumed to be $10^{-3}$ of the true limit on $d_n$.

For case B, the benchmark inputs are: $m_{1/2} = 5.4$ TeV, $m_0 = 12.5$ TeV, $A_0 = 0$, and $\tan \beta = 3.8$. In this case, we see that the current experimental limit on $d_e$ restricts $\theta_\mu/\pi < 0.0048$. And we also see that only if the limit on $d_n$ can be improved by a factor of $\sim 10^3$, can competitive limits on the phase of $\mu$ be obtained. 

\begin{figure}[ht!]
    \centering
    \includegraphics[width=7cm]{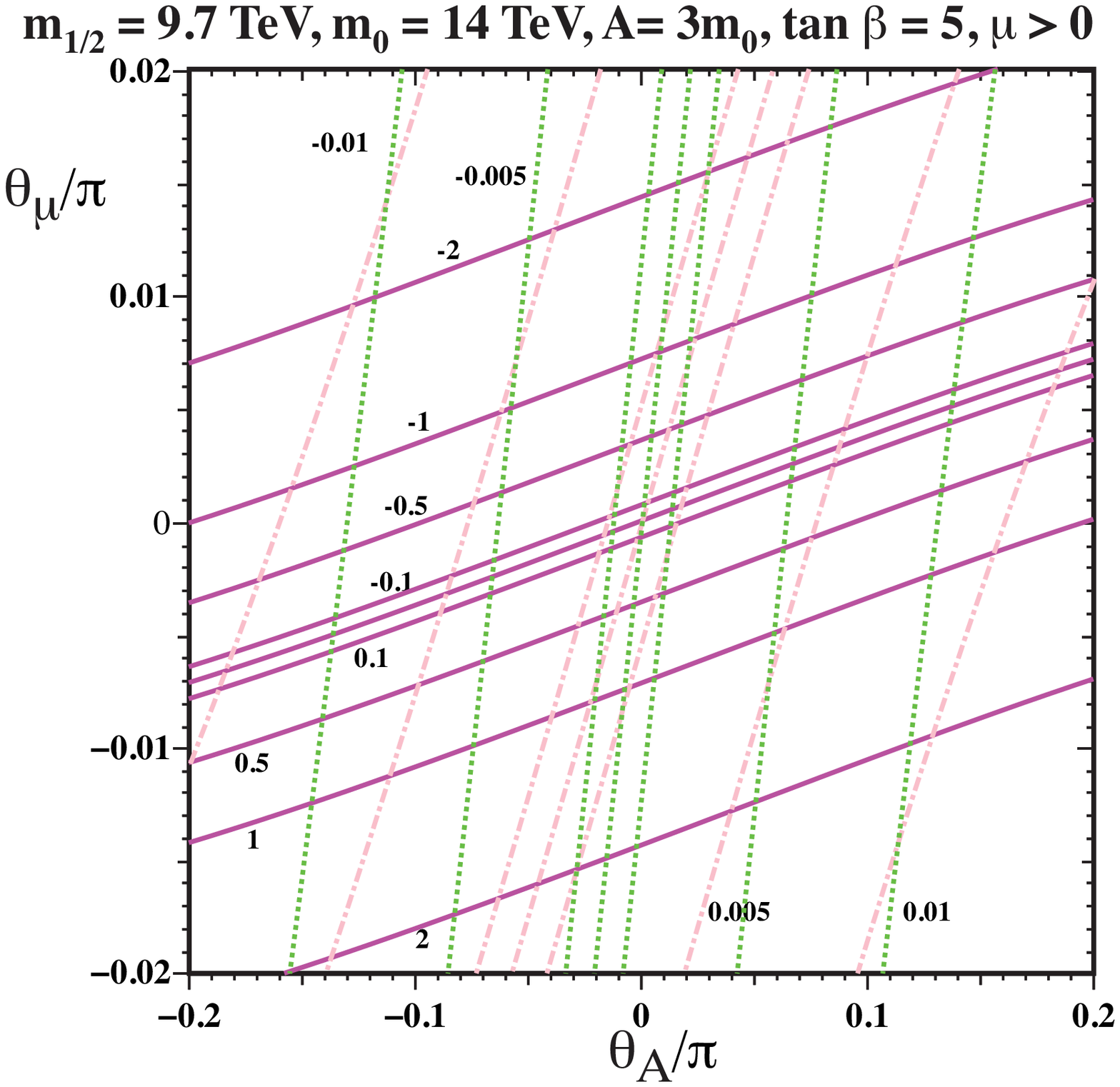}
     \includegraphics[width=7cm]{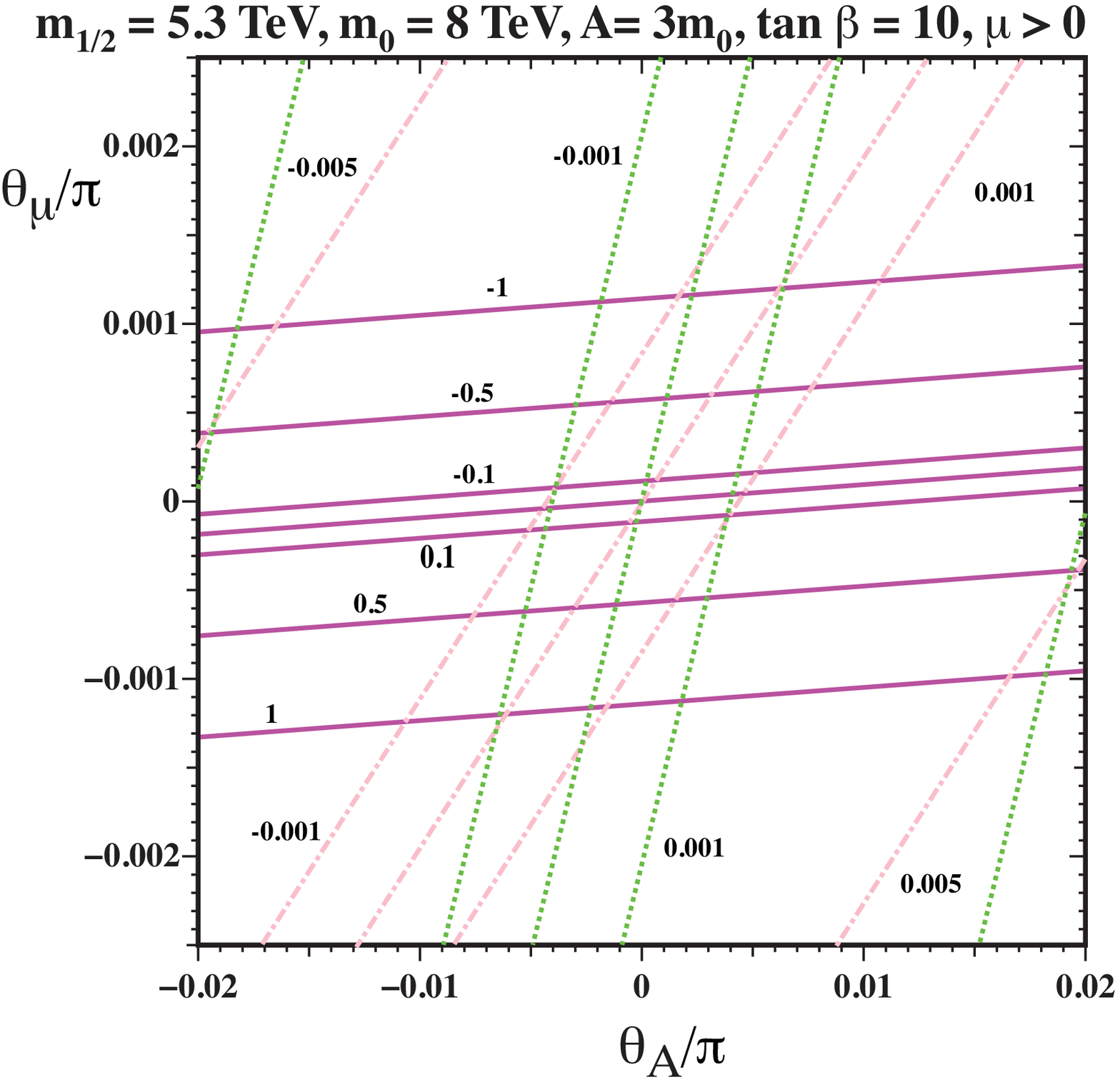} 
    \caption{Contours of the $e$ (solid magenta), $p$ (dotted pink), and $n$ (dot dashed green) EDMs (relative to the experimental limits) in the ($\theta_\mu$, $\theta_A$ plane for fixed $m_{1/2}, m_0, A_0$ and $\tan \beta$ as labelled. For $d_n$  the ratio shown is with respect to $10^{-3}$ times the experimental limit and for $d_p$ with respect to $10^{-29}~e \cdot \mathrm{cm}$. For case C, the unlabeled lines for $d_n$ and $d_p$ are $\pm 0.001$. The central line for each of the EDMs is 0 in both cases C and D. }
    \label{fig:thmuA}
\end{figure}

For cases, C and D, because $A_0 \ne 0$, we can vary both 
$\theta_\mu$ and $\theta_A$.
Plots for two benchmarks in these cases are shown in Fig.~\ref{fig:thmuA}. For case C, the benchmark chosen has 
$m_{1/2} = 9.69$ TeV and $m_0 = 14$ TeV with $A_0 = 3 m_0$ and $\tan \beta = 5$. The calculated mass of the Higgs boson is 123.9 GeV. For case D, 
$m_{1/2} = 5.275$ TeV and $m_0 = 8$ TeV with $A_0 = 3 m_0$ and $\tan \beta = 5$. In this case, $m_H = 125.1$ GeV. Both have a relic density $\Omega_\chi h^2 \approx 0.12$. Through a particular combination of phases,
it is possible to cancel of the phases of each of the EDMs (though simultaneously only at $\theta_A = \theta_\mu = 0$. These are the middle (unlabeled) lines for each type in both panels. The labels correspond to 
the ratio of the calculated EDM with respect to the experimental limit for $d_e$ and with respect to $10^{-3}$ times the limit for $d_n$ and we assume the limit for $d_p$ is $10^{-29}~e \cdot \mathrm{cm}$.

%%%%%%%%%%%%%%%%%%%%%%
\section{Conclusion}
%%%%%%%%%%%%%%%%%%%%%%
Supersymmetric extensions of the Standard Model remain strong candidates for physics beyond the Standard Model. However, if the mass scales associated with supersymmetry breaking are as high as $\mathcal{O}(10)$ TeV, as in the parameter spaces discussed here, the discovery of any supersymmetric particles at the LHC will be difficult. Nevertheless,  there are still ways to probe supersymmetry at a few TeV and even  at $\mathcal{O}(10)$ TeV. These probes prominently include proton decay and CP violating EDMs. 

The aim of this study was to determine if there are
observable effects on the EDMs of the electron, neutron and proton,
despite the large mass scales associated with supersymmetry breaking (due predominantly to the relatively large Higgs mass at 125 GeV). However, 
given the impressive reduction of experimental EDMs limits over the last decade, it is indeed  possible to constrain the CP violating phases entering into the computation of EDMs. In this sense, the SUSY CP-Problem remains present even at $\mathcal{O}(10)$ TeV. 

To show this,  in Sec.~\ref{sec:eff_int}, we first described the formulae that relate the CP-odd interactions in Eq.~\eqref{eq:leff} with the experimentally measured quantities, such as the EDMs of paramagnetic systems and nucleons. We updated these formulae from those used in the previous work~\cite{Demir:2003js,Lebedev:2004va,Olive:2005ru}; in particular, we now partly use the results obtained with QCD lattice simulations to evaluate the nucleon EDMs. In addition, we summarized the current limits on the electron and nucleon EDMs, and discussed future prospects for the measurement of these quantities. The one-loop contribution from supersymmetric particles to the EDMs and CEDMs was given in Sec.~\ref{susyedm} and the higher-order contributions included in our analysis are shown in Appendix.  

Then, in Sec.~\ref{results}, we calculated the electron and nucleon EDMs in the CMSSM. 
 For the cases we have analyzed,  $d_e$ is below the current $\mathcal{O}(10^{-30})~e \cdot \mathrm{cm}$ bound for low $\tan\beta \lesssim 5$, with $A=0$ for a phase $\theta_\mu\leq \mathcal{O}(10^{-1}) \pi$ in the region of compatibility with $m_H\sim 125$ GeV, which is found when $m_{1/2}\sim 40 - 50$ TeV. As we know, important contributions to the EDMs, as well as the predicted mass of the Higgs boson, are enhanced with $\tan\beta$. As a result, for a larger $\tan \beta$, e.g. $\tan \beta = 10$, $m_H = 125$ GeV requires only a gaugino mass of $m_{1/2} \sim 7 - 8$ TeV, and the limit on $\theta_\mu$ improves by an order of magnitude, $\theta_\mu \lesssim 0.001 \pi$. This is a rather surprising result. Even for mass scales of order $\mathcal{O}(10)$ TeV, we are able to set strong constraints on the CP violating phases in the this simple supersymmetric model.  This is entirely due to the improvement in the experimental limits on $d_e$.

Similarly, for cases with $A\neq 0$ we find constraints on the phase of $A_0$,  $\theta_A\leq \mathcal{O}(10^{-1})$ with $\tan \beta = 5$ which is compatible with $m_H\sim 125$ GeV for $m_{1/2} \gtrsim 10$ TeV. For $\tan\beta=10$ the limit improves but is sensitive to $m_{1/2}$. The limit on the phase of $A_0$ in this case is $\theta_A \leq 0.025 \pi$ (for $m_{1/2}\sim 3$ TeV) but relaxes to  $\theta_A \leq 0.1 \pi$ (for $m_{1/2}\sim 5$ TeV). This is consistent with previous work which typically produced weaker bounds on $\theta_A$ than $\theta_\mu$.

As we mentioned in Section \ref{paramag}, there are several prospects for the improvement of the limit on the electron \cite{Alarcon:2022ero}. One of them is the use of new trapping technologies with a potential reach for the EDM of about $10^{-32}\ e\cdot \text{cm}$.  Another is the use of advanced quantum control systems, with a potential reach of about $10^{-34} \ e\cdot \text{cm}$. Using the following scaling for the electron EDM, $d_e$, for cases A and B, in \Figref{fig:thmu}, 
\begin{align}
d_e^A\left(\theta_\mu\right)& =2.2 \times 10^{-29}\ \frac{\theta_\mu}{\pi}\ e\cdot \text{cm}, \nonumber\\
d_e^B\left(\theta_\mu\right)& =2.4 \times 10^{-27}\ \frac{\theta_\mu}{\pi} \ e\cdot \text{cm},
\end{align}
we can estimate when the different planned electron EDM limits will be saturated. For example that for case A, a limit of $1 \times 10^{-34}\ e\cdot \text{cm}$ will be reached for a phase $\theta_\mu/\pi = 5.0 \times 10^{-6}$, while for case B the limit is saturated for a phase $\theta_\mu/\pi = 4.2 \times 10^{-8}$.
For the cases presented in \Figref{fig:thmuA}, we instead give the lines that saturate the limits of $10^{-32}$ and $10^{-34}$. For  case C in \Figref{fig:thmuA}
\bea
\frac{\theta_\mu}{\pi} &=& \pm 1.1 \times 10^{-5} + 4.0 \times 10^{-2}\ \frac{\theta_A}{\pi},\quad \text{for}\ |\theta_A|<\pi \ \times 10^{-3} \quad \text{for}\ \mp 10^{-32} \ e\cdot\text{cm},\nonumber \\
\frac{\theta_\mu}{\pi} &=& \pm 1.1 \times 10^{-7} + 4.0 \times 10^{-2}\ \frac{\theta_A}{\pi}, \quad \text{for}\ |\theta_A|<\pi \ \times 10^{-4} \quad \text{for}\ \mp 10^{-34} \ e\cdot\text{cm}.
\eea
For case D in \Figref{fig:thmuA}, we have
\bea
\frac{\theta_\mu}{\pi} &=& \pm 5.5 \times 10^{-7} + 1.5 \times 10^{-2}\ \frac{\theta_A}{\pi},\quad \text{for}\ |\theta_A|<\pi \ \times 10^{-2} \quad \text{for}\  \mp 10^{-32} \ e\cdot\text{cm},\nonumber \\
\frac{\theta_\mu}{\pi} &=& \pm 5.5 \times 10^{-9} + 1.5 \times 10^{-2}\ \frac{\theta_A}{\pi},\quad \text{for}\ |\theta_A|<\pi \ \times 10^{-3} \quad \text{for}\ \mp 10^{-34} \ e\cdot\text{cm}.
\eea
Evidently, all of these limits roughly follow the simple relation given in \eq{eq:deestimate}, which can be further simplified to 
\beq
|d_e|/(e\cdot {\rm cm})\sim 10^{-28} (10\  \text{TeV}/\tilde{m})^2|\tan\beta\ \theta_\mu -\theta_A| \, ,
\eeq
given the small phases required. However it is important to determine 
the value of $\tilde{m}$ required to be compatible with $m_H \sim 125$ GeV
in each case.

Although we have considered the CMSSM, we expect that in other supersymmetric scenarios, such as an $SU(5)$ no-scale super-GUT model \cite{Ellis:2019fwf}, the current EDM bound can be satisfied. In fact, in \cite{Ellis:2019fwf} it was assumed that the source of CP violation comes only from CKM mixing, fixing it to the observed values at the EW and obtaining as a consequence predictions for other flavor and CP observables, such as EDMs.  Adjusting our predictions from \cite{Ellis:2019fwf} to similar supersymmetry breaking mass scales, the quantity corresponding to $|\tan\beta \ \theta_\mu- \theta_A|$ would be $\sim (10^{-7} - 10^{-6})$, depending on the model. Therefore the prediction for $|d_e|$ is safely below the current limits derived here. 

We have also argued that future improvements of $d_n$ and $d_p$ of $\mathcal{O}(10^{-3})$ have the potential to be competitive with the current limit on $d_e$ in order to constrain the phases. This is particularly relevant for cases where $\theta_\mu$ and $\theta_A$ are both present (likely to be a more realistic scenario) because the shape of the constraints derived from $d_e$, $d_n$ and $d_p$ are all different (see the different slopes in the plots of \Figref{fig:thmuA}) and hence they can really simultaneously corner the values of $\theta_\mu$ and $\theta_A$.

Finally, we remark on the prospects for the EDMs of diamagnetic atoms. At the time of Ref.~\cite{Olive:2005ru}, the limit on the mercury EDM, $|d_{\mathrm{Hg}}| < 2 \times 10^{-28} ~e \cdot \mathrm{cm}$~\cite{Romalis:2000mg}, was used to constrain the parameter space in the CMSSM. This constraint tended to be weaker than those from the neutron and electron EDMs by a factor of $\sim 2$ and $\sim 10$, respectively. Now the mercury EDM bound is improved by a factor of $\sim 30$ ($|d_{\mathrm{Hg}}| < 7.4 \times 10^{-30} ~e \cdot \mathrm{cm}$~\cite{Graner:2016ses}), whilst the neutron and electron EDM limits are improved by factors of $\sim 3$ and $\sim 400$, respectively. We therefore find that for the EDMs of diamagnetic atoms to give limits on the CMSSM competitive to the current electron EDM limit, the sensitivity of their measurement must be improved from the latest mercury EDM measurement by a factor of at least $10^2$.

%%%%%%%%%%%%%%%%%%%%%%%%%%%%%
\section*{Acknowledgements}
%%%%%%%%%%%%%%%%%%%%%%%%%%%%%

The work of N.N. was supported by the Grant-in-Aid for Scientific Research B (No.20H01897), Young Scientists (No.21K13916), and Innovative Areas (No.18H05542).  L. V.-S. acknowledges the support from the ``Fundamental Research Program"  of the Korea Institute for Advanced Study and the warm hospitality and stimulating environment during some stages of the development of this work. The research of L. V-S was supported by the National Research Foundation of Korea (NRF) funded by the Ministry of Education through the Center
for Quantum Space Time (CQUeST) with Grant No. 2020R1A6A1A03047877.  The work of K.A.O.~ and M.P.~was supported in part by DOE grant DE-SC0011842 at the University of Minnesota. 
The work of K.K. was in part supported by the JSPS KAKENHI Grant Nos. 19H01899.

\appendix
\section{Mixing and RG Evolution of Operators
\label{sec:rge}}

The dimension-five (EDM and Chromo-EDM) and dimension-six (Weinberg) operators  in \eq{eq:leff} mix with each other under the renormalization group (RG) flow.  The corresponding Wilson Coefficients, $d_q$, $\ \tilde d_q$ $w$,  at a scale $\mu$ are obtained from those at a higher scale $\mu_0$ from
\bea
\label{eq:mixRGE_edmops}
d_q (\mu)&=&   \eta^{\frac{\gamma_e+\gamma_m}{2 \beta_0}}\, d_q (\mu_S)+  \frac{\gamma_{qe}+\gamma_m}{\gamma_e-\gamma_q} \left[\eta^{\frac{\gamma_e+\gamma_m}{2 \beta_0}} - \eta^{\frac{\gamma_q+\gamma_m}{2 \beta_0}}    \right]  \frac{\tilde{d}_q(\mu_S)}{g_s(\mu_0)}\nonumber\\
& + & \left[a_2 \, \eta^{\frac{\gamma_e+\gamma_m}{2 \beta_0}} + a_3\, \eta^{\frac{\gamma_q+\gamma_m}{2 \beta_0}}  + a_4 \, \eta^{\frac{\gamma_G}{2 \beta_0}} \right] \frac{w(\mu_S)}{g_s(\mu_0)},\\
\tilde{d}_q(\mu) & = &     \eta^{\frac{\gamma_q+\gamma_m}{2 \beta_0}}  \tilde{d}_q(\mu_S) + \frac{\gamma_{Gq}}{\gamma_q+\gamma_m-\gamma_G} \left[\eta^{\frac{\gamma_q+\gamma_m}{2 \beta_0}} - \eta^{\frac{\gamma_G}{2 \beta_0}}    \right] \frac{w(\mu_S)}{g_S(\mu_0)},\\
w(\mu) &=&  \eta^{\frac{\gamma_G-\beta_0}{2 \beta_0}}\, w(\mu_S),
\eea
where the coefficients $a_i$, written in terms of the anomalous dimension matrix, are
\bea
a_2 &=& \frac{\gamma_{Gq} (\gamma_{qe}+\gamma_m) }{(\gamma_q-\gamma_e)(\gamma_G-(\gamma_e+\gamma_m))}, \\
a_3 &=& \frac{\gamma_{Gq} (\gamma_{qe}+\gamma_m) }{(\gamma_e-\gamma_q)(\gamma_G-(\gamma_q+\gamma_m))}, \\
a_4 &=& \frac{\gamma_{Gq} (\gamma_{qe}+\gamma_m) }{(\gamma_e+\gamma_m-\gamma_G)(\gamma_q+\gamma_m-\gamma_G)}.
\eea
We follow \cite{Braaten:1990gq} but notice that our choice of basis is different from all of the cited references and therefore our anomalous dimension matrix elements are different.  In our basis, these are given by
\bea
\gamma_e &=&  8\, C_F,\nonumber \\
\gamma_q &=& 16\, C_F -4N,\nonumber \\
\gamma_{qe} &=& 8\, C_F, \nonumber \\
\gamma_m &=& - 6\, C_F,\nonumber \\
\gamma_{Gq} &=&  -2\, N,\nonumber \\
\gamma_{G} &=&  N + 2\, n_f +\beta_0,\nonumber \\
\beta_0 &=& \frac{1}{3} (11\, N -2\, n_f),
\eea
where $C_F= 4/3$, $N$ and $n_f$ respectively the number of colors and active fermions. At LO the Weinberg operator is zero, but at NLO its main contribution comes from the 2-loop diagram involving the gluino-top-stop contribution. At the scale $\mu_S$, $w^{(1)}$  becomes
\bea
w^{(1)}_{\tilde g}(\mu_S)=-3 \alpha_S \, m_t \frac{g_S}{4\pi} {\rm{Im}} [S^*_{t12} S_{t22}] \frac{z_1-z_2}{m^3_{\tilde g}} H(z_1,z_2,z_t),
\eea
were $S_t$ are the matrices diagonalizing the mass matrix of the s-tops, as defined in \eq{eq:Sq}, $z_\alpha=m_{\tilde \alpha}^2/m_{\tilde g}^2$, $z_t=m_t^2/m_{\tilde g}^2$. 
We first evolve from $\mu_S$, the supersymmetry scale, down to $\mu_5$ where five flavours are active and then from $\mu_5$ down to $\mu_4$, where four flavours are active. At each step we use \eq{eq:mixRGE_edmops}, first with $\mu_0=\mu_S$, $\mu=\mu_5$  and then with  $\mu_0=\mu_5$ and $\mu=\mu_4$.

When the evolution down to a four-flavour theory is taken, we must also consider the shift in $w^{(1)}$ induced by the Chromo-EDM operator. We therefore denote $m_b^+=
\mu_4$ when we take the operators evolved from $\mu_0$ with \eq{eq:mixRGE_edmops} and with $m_b^-=\mu_4$ when we evolve from $\mu_4$ down to the theory with four active flavours. The shift in the Weinberg operator induced by the Chromo-EDM operator can then be written as
\bea
w^{(1)}_{\tilde g}(m_b^-)=
w^{(1)}_{\tilde g}(m_b^+) + g_s(m_b)\, \frac{\alpha_s(m_b)}{8\pi}\,  \tilde d^{\, b\, (0)}_{\tilde g}(m_b),
\eea
where
\bea
w^{(1)}_{\tilde g}(m_b^+) &=& \eta_b^{\frac{39}{46}}\, w^{(1)}_{\tilde g} (\mu_S),\nonumber\\
\tilde d^{\, b\, (0)}_{\tilde g}(m_b)&=&\eta_b^{\frac{14}{23}}\, \tilde d^{\, b\, (0)}_{\tilde g}(\mu_S).
\eea

\section{Notation and two loop contributions to EDM operators
\label{sec:2lsusy}}
We begin with establishing some basic notation. The matrices diagonalizing the sfermions are denoted by $S$, 
\bea
\label{eq:Sq}
S_f^\dagger M^2_{\tilde f}S_f={\rm{diag.}}\left(m^2_{\tilde f_1},\, m^2_{\tilde f_2}\right),
\eea
for $m^2_{\tilde f_1} < m^2_{\tilde f_2}$. For the neutralinos, we choose the convention $\ 
{\mathcal{L}}=-\frac{1}{2}\widetilde\psi^0
{\mathcal{M}}_{\widetilde\psi^0} (\widetilde \psi^0)^T\ $ 
$+\ \quad {\mathrm{h.c.}} \;$,
where the gauge eigenstates are 
$\widetilde\psi^0 = (-i\tilde b, -i\widetilde w, \tilde h_d^0, \tilde h_u^0)$ and the mass matrix is given by
\begin{equation}
\mathcal{M}_{\widetilde\psi^0} = \left[
\begin{array}{cccc} 
M_1 & 0 &
-M_Z\ \cbeta\ \sw & M_Z\ \sbeta\ \sw \\ 0 & M_2 & M_Z\ \cbeta\ \cw &
-M_Z\ \sbeta \ \cw \\ -M_Z\ \cbeta\ \sw & M_Z\ \cbeta \ \cw & 0 & -\mu \\
M_Z\ \sbeta\ \sw & -M_Z\ \sbeta\ \cw & -\mu & 0
\end{array} \right],
\label{eq:mchi0}
\end{equation}
the diagonal matrix is given by $M_{\chi^0}=N^\dagger {\cal M}_{\widetilde\psi^0} N^*$,
and the mass eigenstates, $\tilde\chi^0_i$, are given by $\tilde \chi^0_i=(\tilde\psi^0 N)_{i}$. For charginos, the notation is that of \cite{Haber:1984rc}  with the following identification 
$\tilde\psi^{\pm}=\left(-i \tilde \omega^\pm, \tilde h^\pm_u \right)$, with the Lagrangian $\mathcal{L} =
  -\frac{1}{2}\left[ \widetilde \psi^+ V V^+
 {\cal{M}}_{\widetilde \psi^+} U^* U^T (\widetilde \psi^-)^T + {\rm{h.c.}}\right]$ (where $V$ and $U$ are unitary matrices) and the mass matrix
\begin{equation}
\label{eq:int_basis_charg}
{\cal M}_{\widetilde\psi^+} = \left( \begin{array}{cc} M_2 &
\sqrt2\,M_W\, \cbeta\\\sqrt2\,M_W\, \sbeta & \mu\end{array}\right)
%\label{mchi+}
\end{equation}
which is hence diagonalized by $U$ and $V$, according to  $M_{\rm{diag.}}=M_{\tilde \chi^{\pm}} = V^+ {\cal M}_{\widetilde\psi^+} U^*$.
For the diagonalization of the CP even mass Eigenstates we choose
\bea
{\mathcal{H}}=\left(
\begin{array}{c}
h_1 \\
h_2
\end{array}
\right)=
Z_H \,
\left(
\begin{array}{c}
h \\
H
\end{array}
\right),
\eea
for $h$ and $H$ being respectively the CP neutral mass Eigenstates. 

\begin{figure}
    \begin{center}
    \includegraphics[width=12.5cm]{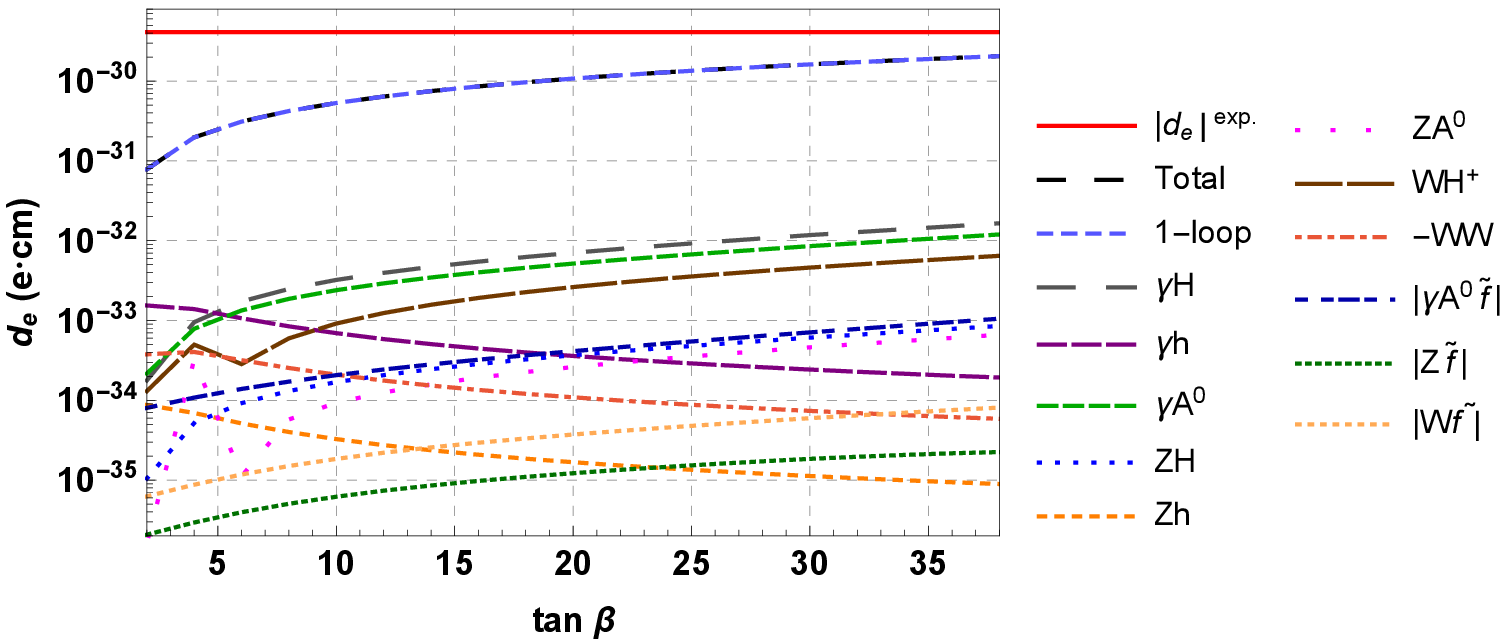}
    \end{center}
    \hspace*{1.88cm}\includegraphics[width=8.5cm]{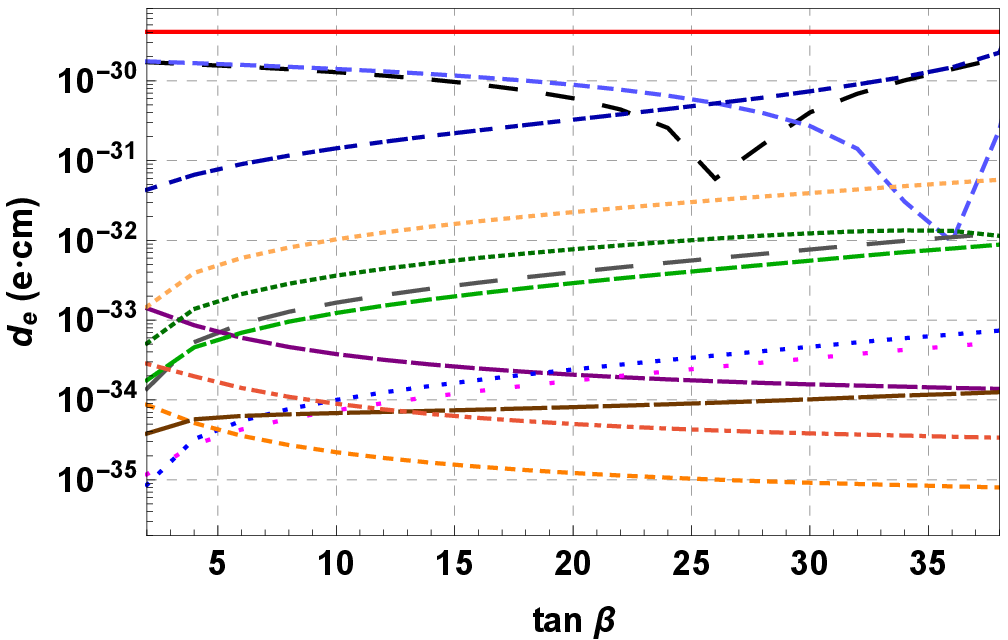}\\
    \hspace*{1.88cm}\includegraphics[width=11cm]{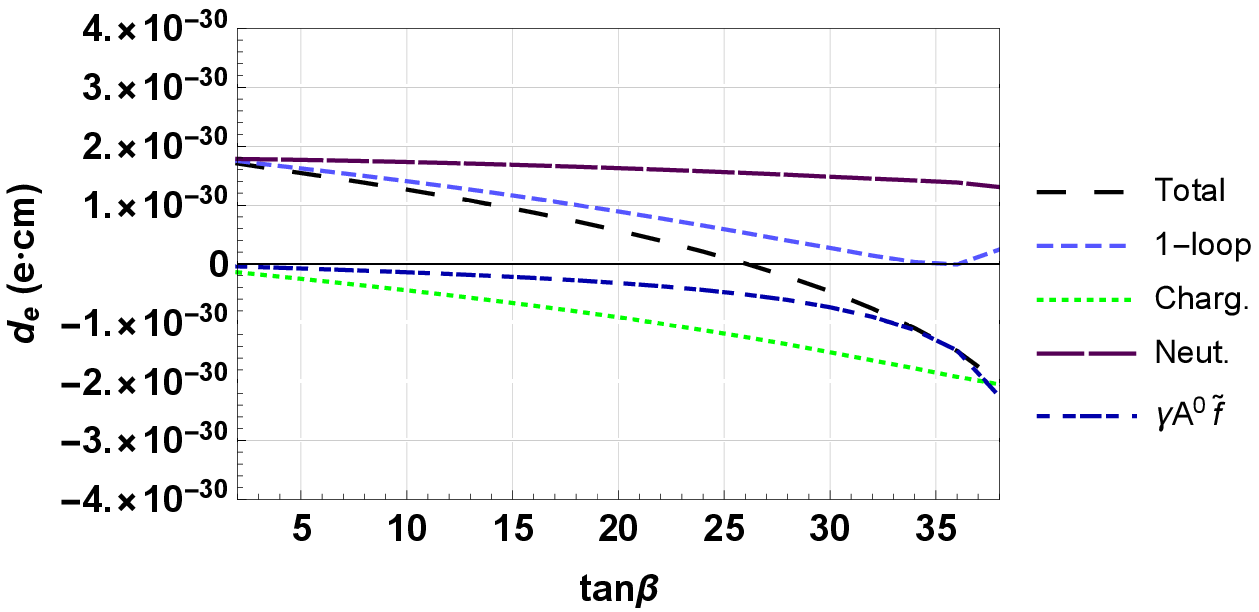}
    \caption{Comparison of the contributions of 1-loop and  2-loop diagrams  as a function of $\tan\beta$ for $\theta_\mu/\pi=0.00015$. $m_{1/2}=8$ TeV and $m_0=8.8$ TeV (upper plot) and $\theta_A/\pi=0.025$ for $m_{1/2}=8$ TeV and $m_0=8.8$ TeV (middle plot). It is well known that the leading 2-loop contribution comes from the diagram $\gamma A^0 \tilde{f}$, specially for large $\tan\beta$ and $\theta_A \neq 0$. To visualize the interplay of the contributions we have also displayed the contributions from charginos, neutralinos and $\gamma A^0 \tilde{f}$ for the case $\theta_A/\pi=0.025$ (lower plot). Basically the contributions from charginos and neutralinos tend to cancel at large $\tan\beta$, according to \eq{eq:deestimate},  and hence when $\tan\beta$ increases the 2-loop contribution from $\gamma A^0 \tilde{f}$ becomes dominant.
    }
    \label{fig:tanb_vs_deEDM}
\end{figure}

\begin{figure}
\begin{center}
\includegraphics[width=14cm]{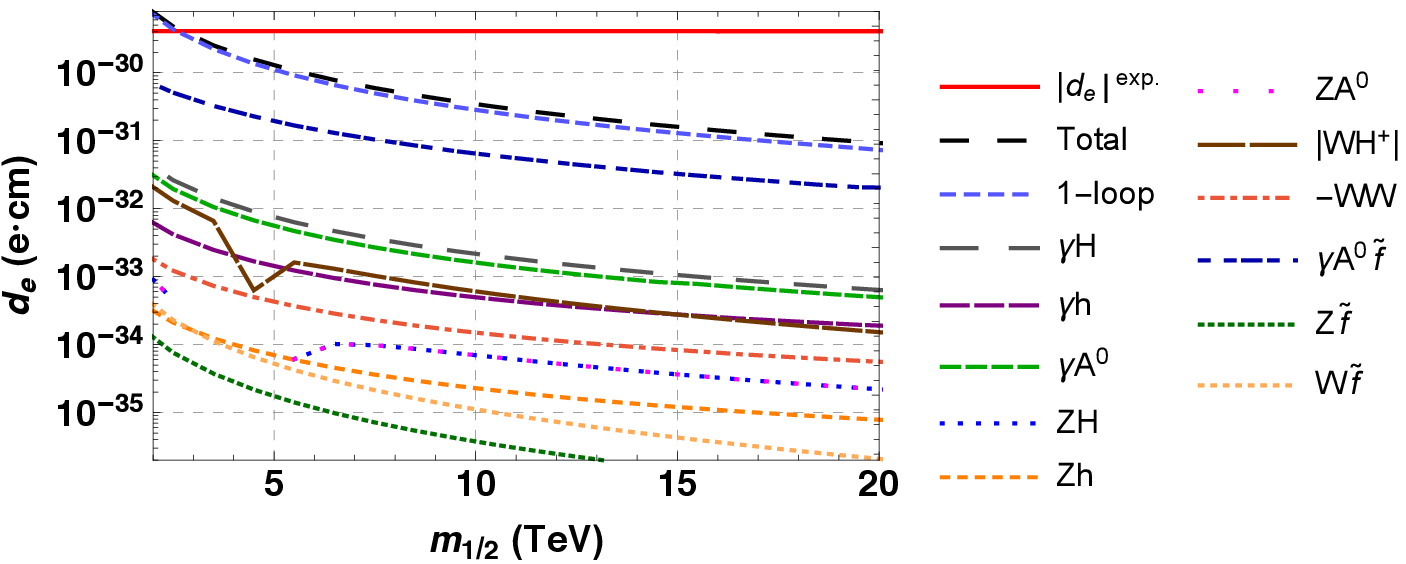}
\end{center}
\hspace*{1.2cm}\includegraphics[width=9cm]{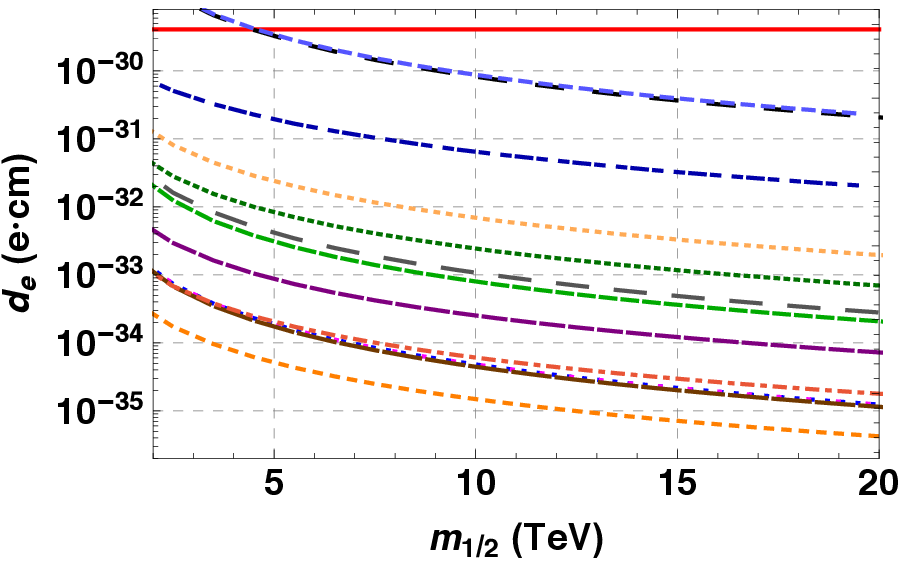}
\caption{Comparison of the contributions from Barr-Zee diagrams as a function of $m_{1/2}$, for $m_0=1.1\times m_{1/2}$, $\tan\beta=10$. Top plot: $\theta_\mu/\pi=0.00015$ and $\theta_A=0$, lower plot: $\theta_\mu=0$, $\theta_A/\pi=0.025$ and $A_0= 3\, m_0$.}
    \label{fig:M12_vs_deEDM}
\end{figure}

The functions entering in the 1-l contributions to the EDMs are as follows
\begin{align}
    A(r) &=
    \frac{1}{2(1-r)^2}\left(3-r+\frac{2\ln r}{1-r}\right),\label{fct:Ar}\\
    B(r) &=
    \frac{1}{2(1-r)^2}\left(1+r+\frac{2r\ln r}{1-r}\right),\label{fct:Br}\\
    C(r) &=
    \frac{1}{6(1-r)^2}\left(10r-26+\frac{2r\ln r}{1-r}-\frac{18\ln r}{1-r}\right).\label{fct:Cr}
\end{align}
At 2-l, the contributions to the EDM from Barr-Zee diagrams are given by 
\bea
d_f^{\gamma\, \mathcal{H}} &=&
 \frac{e\, Q_f \alphem^2 c^{\mathcal{H}}_f} {4\sqrt{2}\, {\pi^2} \swSQ }    \frac{m_f M_i^+}{M_W  m^2_\mathcal{H}}
 \sum^2_{k=1} {\rm{Im}} [D^{\mathcal{H}\, R}_{kk}] f_{\gamma \mathcal{H}} ({r_i^+} _\mathcal{H}), \quad {r_i^+} _\mathcal{H}= \mc^2/\mH^2,\nonumber\\
 && \mH=m_{h},\,  m_{H},
\nonumber\\
d_f^{\gamma\, A^0} &=& \frac{ e\, Q_f \alphem^2 \, c_f^{A^0}}{8 \sqrt{2}\pi^2 \swSQ} \frac{m_f}{M_W \mA^2}  
\sum^2_{i=1} {\rm{Im}} \left[E^R_{ii} \right]\mc
f_{\gamma A^0}(r_{iA^0}), \nonumber\\
d_f^{Z\, \mathcal{H}} &=&
\frac{e\, \alphem^2 (T_{3f} - 2 \swSQ \, Q_f)\, c^{\mathcal{H}}_f} {16 \, \sqrt{2} \, \pi^2  \, \cwSQ\, \swQT}\, \frac{m_f}{M_W\, m^2_{\mathcal{H}}}\,
\sum_{i,j=1}^2\, {\rm{Im}}\, \left[ G^{R}_{ij} D^{R}_{ij} -G^{L}_{ij} D^{L}_{ij}   \right] \mn \nonumber\\
&& \times\ f_{Z \mathcal{H}}(r_{Z\mathcal{H}}, r_{i\mathcal{H}}, r_{j\mathcal{H}}),
\nonumber\\
& & r_{Z\mathcal{H}}=M_Z^2/\mH^2,\ r_{i\mathcal{H}}=\mc^2_{i}/\mcH^2,\
\nonumber\\
d_f^{Z\, A^0} & = & \frac{e\, \alphem^2 (T_{3f} - 2 \swSQ \, Q_f)\, c^{A^0}_f} {16 \, \sqrt{2} \, \pi^2  \, \cwSQ \swQT}\, \frac{m_f}{M_W\, m^2_{A^0}}\,
\sum_{i,j=1}^2\, {\rm{Im}} \left[G_{ij}^R\, E^R_{ji} + G^L_{ij}E^L_{ji} \right] {\mc}_j\nonumber\\ 
& & \times f_{Z\, A^0} \left(r_{Z\, A^0}, r_{i\, A^0}, r_{j\, A^0}  \right), \nonumber\\
 &&  r_{Z\, A^0}=M_Z^2/m^2_{A^0},\,   r_{i,j\, A^0}={\mc}^2_{i,j}/m^2_{A^0} , \nonumber\\
d_f^{W\, W} &=&
\frac{e\, T_{3f} \alphem^2 }{8 \pi^2\, \swQT }{\rm{Im}}\, \sum_{i=1}^{2}\sum_{j=1}^4
\, \left(C^L_{ij}C^{R*}_{ij}\right)\,
\frac{m_f m_i^+ m^0_j}{M_W^4} f_{WW} \left(r^+_{iW}, r^0_{jW} \right),\nonumber \\
&& r^+_{iW}= \left(m_i^+/M_W\right)^2, \, 
r^0_{iW}=\left(m_j^0/M_W\right)^2,
\nonumber\\
d_f^{W\, H^{\pm}} &=&
-\frac{e\, \alphem c^{H^{\pm}}_f }{ 32\ \pi^2 \swQT\, \cw} \frac{m_f}{M_W \mc^2} \nonumber\\
& & \times \ \sum_{i=1}^{2}\sum_{j=1}^4
\, {\rm{Im}} \left[C^L_{ij}{F^L_{ij}}^* + C^R_{ij}{F^R_{ij}}^*  \right]
\mc_i \ f_{1\, W \mcH} (r_{W\mcH}, r_{i\mathcal{H}}, r_{j\mathcal{H}} )\nonumber\\
&&+\ {\rm{Im}} \left[C^R_{ij}{F^L_{ij}}^* + C^L_{ij}{F^R_{ij}}^*  \right] \mn_j
f_{2\, W \mcH} (r_{W\mcH}, r_{i\mathcal{H}}, r_{j\mathcal{H}} )
\nonumber\\
&& 
+ \ {\rm{Im}} \left[C^L_{ij}{F^L_{ij}}^* - C^R_{ij}{F^R_{ij}}^*  \right] \mc_i 
f_{3\, W \mcH} (r_{W\mcH}, r_{i\mathcal{H}}, r_{j\mathcal{H}} )
\nonumber\\
&& +\ {\rm{Im}} \left[C^R_{ij}{F^L_{ij}}^* - C^L_{ij}{F^R_{ij}}^*  \right] \mn_j
f_{4\, W \mcH} (r_{W\mcH}, r_{i\mathcal{H}}, r_{j\mathcal{H}} ),
\nonumber\\
& & r_{W\mathcal{H}}=m^2_W/\mH^2,\ r_{i\mathcal{H}}=\mc^2/\mcH^2,\  r_{j\mathcal{H}}=\mn^2/\mcH^2.
\eea
For the couplings involving Higgs particles and sfermions we have\footnote{It is well known that the leading 2-loop contribution comes from the diagram $\gamma A^0 \tilde{f}$, specially for large $\tan\beta$ and $\theta_A \neq 0$.} \cite{Pilaftsis:1999td}
\begin{eqnarray} 
  \label{eq:HiggsGZW}   
  d_f^{\, \gamma A^0 \tilde{f}} &=&   
\frac{e\, Q_f  \alphem\, N_c}{32\pi^3}\, \frac{\tan\beta\ m_f}{m^2_{A^0}}\ \sum_{f' = t,b,\tau}\ 
c_{\tilde{f'} A^0}\, Q^2_{f'}\,\left[\, F\left(r_{\tilde{f'}_1 A^0}\right)\ -\ 
F\left(r_{\tilde{f'}_2 A^0}\right)\, \right]\, , \nonumber\\
& &  r_{\tilde{f'}_i A^0}=m^2_{\tilde{f'}_i}/{m^2_{A^0}},\
N_c, \ {\small{{\rm is\ the\  color}\ {\rm factor}}}\,
,\nonumber\\
d_f^{\, Z \tilde{f}} &=& -e\, (T_{3f} - 2Q_f \swSQ)\,   \frac{\alphem\, N_c}{64 \, \swSQ\cwSQ\, \pi^3}\, 
\frac{\tan\beta\ m_f}{M^2_{A^0}}\nonumber\\
&&\times \sum_{f' = t,b,\tau}\ 
\sum_{i,j=1,2}\   
c_{A^0\tilde{f}_i^{\prime *}\tilde{f'}_j}\, K_{f' ij}\, Q_{f'}\
G\left( r_{Z A^0},\ r_{\tilde{f'}_i A^0},\ 
r_{\tilde{f'}_j A^0}\right)\, ,\nonumber \\ 
&& r_{Z A^0}=M_Z^2/{m^2_{A^0}},\nonumber\\
d_f^{\, W \tilde{f}} &=&    
\frac{e\, N_c \alphem}{128\, \swSQ\, \sqrt{2} \pi^3}\, 
\frac{\tan\beta\ m_f}{M^2_{H^+}}\,  
\sum_{i,j=1,2}\ c_{H^+ \tilde{t} \tilde{b}}\, K^{tb}_{ij}\nonumber\\
&&\times\,
\left[\,Q_t\, G\left(r_{W H^+},\ 
r_{\tilde{t}_i H^+},\ 
r_{\tilde{b}_j H^+}\right)\,
+\, Q_b\, G\left(
r_{W H^+},\ 
r_{\tilde{b}_i H^+},\ 
r_{\tilde{t}_j H^+}
\right)\, \right]\,,\nonumber\\
&& r_{W H^+}=M_W^2/{m^2_{H^+}},\
r_{\tilde{q}_i H^+}=m_{\tilde{q_i}}^2/{m^2_{H^+}}, q=t,b.
\, \nonumber\\
\end{eqnarray}
All the couplings in the expression above are given by
\bea
G^L_{ij}&=&\frac{1}{2}\left( V_{1i} V^*_{1j} + \delta_{ij}(\cwSQ-\swSQ) \right),\nonumber\\
G^R_{ij}&=&\frac{1}{2}\left( U_{1i}^* U_{1j} + \delta_{ij}(\cwSQ-\swSQ) \right),\nonumber\\
C^{L}_{ij} &=& V_{1i}\, N_{2j} - \frac{1}{\sqrt{2}} V_{2i}\, N_{4j},\nonumber\\
C^{R}_{ij} &=& U^*_{1i}\, N_{2j}^* + \frac{1}{\sqrt{2}} U^*_{2i}\, N_{3j}^*,\nonumber\\
D^{\mathcal{H}\, R}_{ij} &=& (Z_H)_{1k}\, V_{1i} U_{2j} + (Z_H)_{2k}\,
V_{2i}  U_{1j},\ k=1 \ {\text{for}}\ h, \ k=2 \ {\text{for}}\ H,\nonumber\\
D^{\mathcal{H}\, L}_{ij} &=& (Z_H)_{1k}\, V_{1j}^* U_{2i}^* + (Z_H)_{2k}\,
V_{2j}^* U_{1i}^*, \ k=1 \ {\text{for}}\ h, \ k=2 \ {\text{for}}\ H, \nonumber\\
E^{ L}_{ij} &=& -\sbeta\, U^*_{2i} V^*_{1j}  - \cbeta\,\, U^*_{1i} V^*_{2j},\nonumber\\
E^{ R}_{ij} &=& \sbeta\, U_{2j} V_{1i}  + \cbeta\, U_{1j} V_{2i},\\
F^L_{ij} &=&\sbeta\left(\frac{1}{\sqrt{2}}\, U^*_{2i} (N_{1j} \sin \theta_{\mathrm{W}} + N_{2j} \cw)  - U_{1i} N_{3j}\cw  \right),\nonumber\\
F^R_{ij} &=& -\cbeta \left( \frac{1}{\sqrt{2}}
V_{2i} \left( N_{1j}^* \sw + N_{2j}^* \cw\right)
+ V_{1i} N_{4j}^* \cw
\right).
\eea
The coefficients $c_f^{h,H,A^0, H^{\pm}}$ are as follows
\begin{align}
&c_d^h=c_e^h=\frac{{(Z_{H})}_{11}}{\cos\beta},\, \quad c_u^h=\frac{{(Z_{H})}_{21}}{\sin\beta},\quad
c_d^H=c_e^H=\frac{{(Z_H)}_{12}}{\cos\beta},\nonumber\\
&c_d^{A^0}=c_e^{A^0}=\tan\beta,\ c_u^{A^0}=\cot\beta,\
c_d^{H^+}=c_e^{H^+}=\tan\beta,\
c_u^{H^+}=\cot\beta.
\end{align}
The remaining quantities are as follows
\bea
c_{\tilde{f} A^0} &=& 2\, \tan\beta^p\, \frac{ |S_{q12}||S_{q22}|\, m_q\, {\rm Im}[\mu e^{i \delta_q }] }{v^2\, \sin\beta \cos\beta }\, , \nonumber\\
\delta_f &=&-{\rm arg}(A_f+\tan\beta^p\mu^*),
 \ p=1 \ {\rm for}\  f=b,\tau, \ p=-1 \   {\rm for} \ f=t,\,
\nonumber\\
c_{A^0\tilde{f}_i^*\tilde{f}_j}&=& c_{\tilde{f} A^0}  \, \left(
\begin{array}{cc}
1 & -\frac{|S_{f22}|^2-|S_{f12}|^2}{2 |S_{f22}|^2}\\
-\frac{|S_{f22}|^2-|S_{f12}|^2}{2 |S_{f22}|^2} & 1
\end{array}
\right),\, \nonumber \\
K_f &=& \left(
\begin{array}{cc}
 2 T_{3f} |S_{f22}|^2-2 Q_f\ \swSQ     & 2\, T_{3f} |S_{f12}||S_{f22}|\\
 2\, T_{3f} |S_{f12}| |S_{f22}|    & 
  2 T_{3f} |S_{f12}|^2-2 Q_f\ \swSQ
\end{array}
\right)\, ,\nonumber\\
K^{tb} &=& 
\left(\begin{array}{cc}
|S_{t22}||S_{b22}|  & |S_{t22}||S_{b12}|\\
|S_{b22}||S_{t12}|  & |S_{t12}||S_{b12}|
\end{array}
\right)
,\nonumber\\
c_{H^+ \tilde{t} \tilde{b}}&=& \frac{1}{\sqrt{2}}
\left(
\begin{array}{cc}
-P_{tb}\, c_{\tilde{b} A^0}+ P_{bt}\, c_{\tilde{t} A^0} & c_{\tilde{b} A^0}/Q_{tb}+ Q_{bt}\, c_{\tilde{t} A^0} \\
-Q_{tb}\, c_{\tilde{b} A^0} - c_{\tilde{t} A^0}/ Q_{bt} &
R_{tb}\, c_{\tilde{b} A^0} - R_{bt}\, c_{\tilde{t} A^0} 
\end{array}
\right) \nonumber\\
&+ & \frac{\sqrt{2}\, m_b m_t \sin(\delta_b-\delta_t) }{v^3\sin\beta \cos\beta }
\left(\begin{array}{ccc}
|S_{b 12}|S_{b 12}| & - |S_{t 12}| | S_{b 22}|\\
-|S_{b 12}| |S_{t 22}| &  |S_{t 22}||S_{b 22}|
\end{array}
\right)\, ,\nonumber\\
&& P_{q_1 q_2} =(S_{q_1})_{22}/(S_{q_2})_{22},\quad
Q_{q_1 q_2}= (S_{q_1})_{12}/(S_{q_2})_{22},\nonumber\\
&& R_{q_1 q_2}=(S_{q_1})_{12}/(S_{q_2})_{12}.
\eea
For clarity of the presentation we give explicitly all of the loop functions, that can also be found in \cite{Barr:1990vd},
\bea
f_{\gamma \mathcal{H}}(r) &=& K_0(r), \nonumber\\
f_{\gamma A^0}(r)&=& K_0(r) -2 K_1(r) +2 K_2(z),\nonumber\\
K_n(r) &=& \int^1_0\, dx\, \frac{x^n}{r-x(1-x)}\, \ln\left(\frac{r}{x(1-x)}\right),\nonumber\\
f_{Z \mathcal{H}} (r_1,r_2,r_3) &=&  \int_0^1\, \frac{1}{x}\, J\left(r_1,\frac{x r_2 + (1-x) r_3}{x\, (1-x)}\right) ,\nonumber\\
f_{ Z\, A^0} (r_1,r_2,r_3)  &=&  \int_0^1\, \frac{1-x}{x}
J(r_1, \frac{x r_2 + (1-x) r_3}{x\, (1-x)}),\nonumber\\
f_{WW} \left(r_1, r_2 \right)
&=& \int^1_0 \frac{dx}{1-x}\, J\left(0, \frac{x r_1 + (1-x)\, r_2}{x (1-x)}\right)
,\nonumber\\
f_{1\, W \mcH} (r_1,r_2,r_3) &=&   \int_0^1\, dx\, \frac{x^2}{1-x} J\left(r_1, \frac{r_2}{1-x} + \frac{r_3}{x} \right), \nonumber\\
f_{2\, W \mcH} (r_1,r_2,r_3) &=& \int_0^1\, dx\, (1+x)\, J\left(r_1, \frac{r_2}{1-x} + \frac{r_3}{x} \right), \nonumber\\
f_{3\, W \mcH} (r_1,r_2,r_3) &=& \int_0^1\, dx\, \frac{x}{1-x}\, J\left(r_1, \frac{r_2}{1-x} + \frac{r_3}{x} \right), \nonumber\\
f_{4\, W \mcH} (r_1,r_2,r_3) &=&  \int_0^1\, dx\, J(r_1, \frac{r_2}{1-x} + \frac{r_3}{x} ), \nonumber\\
J(r) &= & \frac{r\, \ln(r)}{r-1},\nonumber\\
J(r_1,r_2)&=&\frac{ J(r_1) - J(r_2)}{r_1-r_2},\nonumber\\
F(r_1) &=& \int_0^{1} dx\ \frac{x(1-x)}{r_1 - x(1-x)}\ 
\ln \left(\,\frac{x(1-x)}{r_1}\,\right)\, , \nonumber \\
G(r_1,r_2,r_3) &=& \int_0^1 dx\ x\, \left[\, \frac{r_1 x(1-x) \ln r_1}{
(r_1-1) [r_1 x(1-x) - r_2 x - r_3 (1-x)]}\right.\nonumber\\ 
&&\hspace{-2cm}+ \left. \ \frac{ x (1-x) [ r_2 x + r_3 (1-x)]}{
[r_1 x(1-x) - r_2 x - r_3 (1-x)]\, [x(1-x) - r_2 x - r_3 (1-x)] }\right. \times \\ \nonumber
&&\left. \ln \left( \frac{ r_2 x + r_3 (1-x)}{ x (1-x) } \right)\, \right]\, .
\eea
In \Figref{fig:tanb_vs_deEDM} we present a comparison of the 1-loop and 2-loop contributions described here, as a function of $\tan\beta$ for $A=0$, $m_{1/2}=8$ TeV and $m_0=8.8$ TeV for $\theta_\mu/\pi=0.00015$ (upper plot) and $A=3\, m_0$ and  $\theta_A/\pi=0.025$ (middle plot). For this last case we have presented a decomposition of chargino and neutralino contributions to show how these two contributions tend to cancel each other at large $\tan\beta$, making $\gamma A^0 \tilde{f}$ the dominant contribution (lower plot) and even contributions from $W\tilde{f}$, $\gamma H$, $\gamma A^0$ and $Z\tilde f$ become larger than the 1-loop contribution. In \Figref{fig:M12_vs_deEDM} we plot instead the 1-loop vs the 2-loop contributions as a function of $m_{1/2}$ for $\tan\beta=10$, $m_0=1.1 \times m_{1/2}$ for $A=0$ and $\theta_\mu/\pi=0.00015$ (upper plot) and for $A= 3\, m_0$ and $\theta_\mu/\pi=0.025$.  


\end{document}